\documentclass[aps,prb,onecolumn]{revtex4}

\usepackage{bm}
\usepackage{amssymb}
\usepackage{graphicx}
\usepackage{subfigure}
\usepackage{epstopdf}
\usepackage{amsmath}
\usepackage{color}
\usepackage{microtype}
\usepackage{float}
\usepackage[english]{babel}
\usepackage{braket}
\usepackage{comment}

\newcommand{\sfa}{\sf a}
\newcommand{\sfb}{\sf b}

\newcommand{\iu}{\mathrm{i}}
\newcommand{\eu}{\mathrm{e}}
\makeatletter
\newcommand\figcaption{\def\@captype{figure}\caption}
\makeatother

\newcommand{\mbf}[1]{\mathbf{#1}}

\newcommand{\Qp}{Q}

\begin{document}

\title{ Statistical mechanics models
for multimode lasers and random lasers  
} 

%\author{
%name{F. Antenucci \textsuperscript{a,b}$^{\ast}$
%  \thanks{$^\ast$Corresponding author. Email: fabrizio.antenucci@roma1.infn.it}
% A. Crisanti \textsuperscript{b,c}
%  M. Ib\'a\~nez Berganza \textsuperscript{b,d}
% A. Marruzzo \textsuperscript{a,b} 
% and  
% L. Leuzzi \textsuperscript{a,b}
% } 
%\affil{
%\textsuperscript{a} NANOTEC-CNR, Institute of Nanotechnology, Soft and Living Matter Lab, Rome, Piazzale A. Moro 2, I-00185, Roma, Italy
%\textsuperscript{b} Dipartimento di Fisica, Universit\`a di Roma ``Sapienza,'' Piazzale A. Moro 2, I-00185, Roma, Italy
%\textsuperscript{c} ISC-CNR, UOS {\it Sapienza}, Piazzale A. Moro 2, I-00185, Roma, Italy
%\textsuperscript{d} INFN, Gruppo Collegato di Parma, via G.P. Usberti, 7/A - 43124, Parma, Italy
%}
%}

 \author{F. Antenucci $^{1,2}$, A. Crisanti$^{2,3}$, M. Ib\'a\~nez Berganza$^{2,4}$, A. Marruzzo$^{1,2}$, L. Leuzzi$^{1,2}$ }
 \email{fabrizio.antenucci@roma1.infn.it} 
 \affiliation{
 $^1$ NANOTEC-CNR, Institute of Nanotechnology, Soft and Living Matter Lab, Rome, Piazzale
  A. Moro 2, I-00185, Roma, Italy \\
 $^2$ Dipartimento di Fisica,
  Universit\`a di Roma ``Sapienza,'' Piazzale A. Moro 2, I-00185,
  Roma, Italy  \\ $^3$ ISC-CNR, UOS {\it Sapienza},
  Piazzale A. Moro 2, I-00185, Roma, Italy
  \\
  $^4$ INFN, Gruppo Collegato di Parma, via G.P. Usberti, 7/A - 43124, Parma, Italy
  }

\begin{abstract}
We review recent statistical mechanical approaches to %the study of %semiclassical 
multimode laser theory.
The theory has proved very effective to describe 
standard lasers. 
We refer of the mean field theory for passive mode locking
and developments based on Monte Carlo simulations and cavity method
to study the role of the frequency matching condition.
The status for a complete theory of multimode lasing in open and disordered cavities
is discussed and
the derivation of the general statistical models in this framework is presented.
When light is propagating in a disordered medium,
the system can be analyzed via the replica method.
For high degrees of disorder and nonlinearity,
a glassy behavior is expected at the lasing threshold,
providing a suggestive link between glasses %and structural glasses 
and photonics.
We describe in details the results for the general Hamiltonian model in mean field approximation
and mention an available test for replica symmetry breaking
from intensity spectra measurements.
Finally, we summary some perspectives still opened for such approaches.
\end{abstract}

\maketitle

%\begin{keywords}
% statistical mechanics, optics, critical phenomena, lasers, glass transition, phase transitions
%\end{keywords}

The idea that the lasing threshold can be understood as a
proper thermodynamic transition goes back since 
the early development of laser theory and non-linear optics in the 1970s,
in particular in connection with modulation instability
(see, e.g., \cite{Haken78} and the review \cite{AreBocRam99}).
A proper formalization of this idea  
was achieved in the '00s, with the works of B. Fischer et al. \cite{GorFis02,GorFis03,Goretal03,Vodetal04,Weietal05},
where the statistical properties of laser light in homogeneous cavities are
investigated taking into account nonlinear effects, 
like gain saturation and intensity dependent refractive index.
Mapping the laser dynamics in an ordered Hamiltonian problem, 
with the nonlinearities providing the interaction between the electromagnetic modes,
the authors of Ref. \cite{Weietal05} show 
new aspects and reveal, in particular,
that a critical behavior %(a phase transition) 
occurs at the laser mode locking transition.
%{\bf [vedere per referenze pi\'u recenti di Fisher \& c]}
%In this case, the statistical mechanics of ordered 
%systems is applied to study light propagation in
%amplifying homogeneous nonlinear materials. 
In section \ref{sec:ordered} of this paper 
we review the main results and the most recent developments \cite{AntIbaLeu14b,MarLeu15,AntIbaLeu15}
of this use of statistical mechanics of ordered systems %approach 
for the study of light propagation in nonlinear materials in homogeneous cavities.

In the last years the role of local inhomogeneities in matter and
its disordered nature %of these inhomogeneities \red{its disordered nature [senza virgola]} 
has proved to be at the origin of a variety of novel interesting phenomena.
Specifically, 
light amplification in random media and random lasers (RL) 
have attracted much attention
\cite{WieLag96,Cao01,Wiersma01,Anglos04,Sharmaetal06,Mujetal07,Leprietal07,vanderMolen07,Wiersma08,Ghofraniha15}.
This is an intriguing topic that crosses different fields
as light localization, non-linear physics %,thermodynamics,
and quantum optics and has relevant fundamental aspects and as well practical perspectives like, e.g., 
biomedical diagnostics \cite{RedChoCao12}, chip-based spectrometers \cite{RedCao12,RedPopCao13,Redetal13}
and cryptography \cite{Horetal13}. 
%The complex structure and the extreme openness make these optical systems different from traditional cavity laser composed of lossless cavity.
%

From a historical point of view, the presence of a divergence in the intensity for light diffusion with gain
above a critical volume was already discussed by Letokhov in the 1960s.
%In this situation, if the gain depends on the wavelength, the emission spectrum narrows down close to the wavelength of maximum gain.
%These features were later observed in experiments \cite{MarZolBri86,Gouetal93}.
%
However, lasing occurs in general as a result of two basic ingredients: \emph{optical amplification} and \emph{feedback}.
Amplified spontaneous emission % (ASE) 
occurs even without optical cavity, the spectrum is completely determined by the gain curve of the active material.
The material is called a RL only when the multiple-scattering process (feedback) plays a key role in determining the lasing process \cite{WieLag96}.
%The presence of feedback is associated to the existence of well-defined long-living cavity modes characterized by a definite
The presence of feedback is associated to the existence of well-defined cavity modes with a long-life and characterized by a definite
spatial pattern of the electromagnetic field, which sets up inside the lasing structure in the stationary regime. 
%In this case interference effects determine the mode structure. 
%In a regular laser the modes are determined by the laser cavity and consist of the standing-wave pattern. 
%In a RL, on the other hand, it is the multiple-scattering process that defines the optical modes, with a certain frequency, bandwidth and %rich 
%complicated spatial profile. % (cf. Chapter \ref{chapterII}).
%A RL is, in other words, \emph{``mirror-less''} but not \emph{``mode-less''} \cite{wiersma2008physics}.

A basic theory for RLs may, in principle, be developed along the lines of standard semiclassical
lasing theory, but it has to incorporate the specific features of random lasers:
a much larger role of radiative leakage and an irregular spatial structure of the modes.
While this theory is far from being complete, a review of some recent developments,
including the essential elements for a statistical approach,
is reported in section \ref{sec:RL_optics}.
In this physical situation, the methods of statistical mechanics have proved to be very powerful  
to treat the strong interplay between disorder and nonlinearity \cite{Angetal06,Leuetal09,Antetal14}.
%and foreseen that 
Within this approach, 
it is possible to draw suggestive analogies between 
propagation of light in nonlinear disordered media and the rich behavior observed in glasses.
We report some of the most important results from this approach in section \ref{sec:RL_statMec}.

%%%%%%%%%%%%%%%%%%%%%%%%%%%%%%%%%%%%%%%%%%%%%%%%%%%%%%%%%%%%%%%%%%%%%%%%%%%%%%%%
\section{The case for standard lasers}
\label{sec:ordered}

The statistical mechanics paradigm
called Statistical Light-mode Dynamics (SLD)
has been introduced in the early works \cite{GorFis02,GorFis03}
to study multimode laser physics, 
for which the number of modes is generally large enough ($10^2 - 10^9$ in long lasers),
and such that the presence of nonlinearity makes the problem nontrivial.
One of the main benefits of this approach is the recognition of the role of the noise,
due, e.g., to the unavoidable spontaneous emission.
%Note that the dynamics of a laser is always subject to noise:
%besides the typical noise sources always present in physical systems, there is the inevitable fundamental noise of the spontaneous emission. 
%Consequently, a model %for lasers 
%that does not acknowledge noise properly may risk missing crucial features in the laser physics. 
In literature, see, e.g., Refs.  \cite{Ho85,RanRamHau04}, classical and quantum noise is often considered as a small perturbation 
of the noiseless pulse of the laser master equation.
%The SLD approach allows instead to handle the noise, that takes here the role of temperature in the thermodynamic system, nonperturbatively.
In the SLD approach, instead, the noise takes the role of the temperature in the thermodynamic system and 
can be treated nonperturbatively.
As a particular result, it is shown that the perturbative approach is in general not %justified: 
legitimate: accumulation of noise in the whole cavity 
can generate a continuous background that carries a significant portion of the total optical %cavity 
power and competes with the pulse \cite{GorFis02}.

The entropy associated with the noise becomes an essential ingredient in the study of \emph{mode locking},
the regime under which a laser generates ultra-short pulses \cite{GorFis03b}.
%
%Understanding the conditions for mode locking is a theme of great interest, both theoretical and practical. 
The pulse formation in lasers is based on the interaction between axial modes. 
Such an interaction can be provided either by making the system 
%time dependent (modulating) or by a suitable nonlinearity in the dynamics of the system.
time dependent (\emph{active mode locking}) or by a suitable nonlinearity in the dynamics of the system (\emph{passive mode locking}) \cite{Haus75,KuiSie70}.
%
%One type of nonlinearity known to encourage pulsed operation is achieved
%incorporating a saturable absorber into the laser resonator.
%
%In the mode domain the
%saturable absorber induces a nonlinear %four-wave-mixing interaction 
%4-mode interaction.
%between the modes
%The some kind of nonlinearity is induced by the Kerr effect, with the difference that it is dissipative rather than dispersive.
%
%
Within the SLD the difference between active and passive mode locking is clear
and it is embedded in the range of the interaction between modes in the two cases.
The active case corresponds to a one-dimensional short-range-interacting model \cite{GorFis04}: 
in this case a phase transition to an ordered state 
occurs in principle only at zero noise.
%\footnote{ 
%In a finite system 
%this picture is only precise when the spectral correlation length is shorter than the finite bandwidth,
%so that the number of modes involved in the laser dynamic is large enough.
%}
The fragility of active mode locking becomes, hence, an exemplification 
of the well-known lack of \emph{global ordering} %(``magnetization'') % in spin systems) 
of the one-dimensional %short-range-interacting 
spherical spin model \cite{BerKac52}:
any weak noise %in the mode system
can break a bond between two interacting % neighboring 
modes, thus eliminating global mode ordering.
%\footnote{  
% This is exactly true for the usual harmonic modulation. %($\eta = 2$), 
% If instead, the modulation is a not-smooth function of time with a power law singularity, %otherwise ($\eta < 2$) 
% the system undergoes a continuous (unlike the passive mode locking case) phase transition to a Bose-Einstein Condensate (BEC)  
% phase transition \cite{Fisher2010, Fratalocchi10, Fisher_experiment_cond14}.
%}
%
%
In contrast, the long-range interaction in passive mode locking,
due to the four-wave mixing in the saturable absorber, 
imposes a global order below a certain noise level,
resulting in the threshold behavior:
once the interaction %through the many mode ? mode bonds 
is strong enough to overcome noise it induces long-range correlations %in all the system % all over the band. 
and a first order phase transition between disordered (continuous wave) and ordered (mode locked)
thermodynamic phases \cite{GorFis02,GorFis03,GatGorFis04,Vodetal04}. %\red{[se vuoi, la maledetta tabella qua?]}
Thus, the individual action of the noise stabilizes the continuous wave regime,
showing a noise induced phase transition \cite{VanParTor94}. 

Several %Many 
other theoretical and experimental features can be studied in SLD. 
%among them: hysteresis, 
%superheating and supercooling, successive formation of multiple pulses in the cavity and more.
Among the others, 
we mention the addition of an external driving field to a passive mode locked system, 
analog to the external magnetic field in magnets or to the pressure in gas-liquid-solid systems,
representing the injection in the laser cavity of pulses from an external source,
which in the simple case matches the repetition rate of the laser \cite{Weietal05}. 
%When the injection is weak, the ordering phase transition persists, shifted to higher ``temperature''.
In this situation, beyond a threshold injection level, the transition becomes continuous
rather than first order. The two phase transition lines meet
at a tricritical point around which tricritical behavior is observed.

For a passive mode locking laser, the SLD description can be obtained considering
the standard master equation \cite{Haus00}
\begin{align}
 \frac{d a_l}{dt}  = 
 \left( G_l + \iu D_l \right) a_l (t)
 + \left( \Gamma - \iu \Delta \right) 
   \sum_ { k_1-k_2+k_3-l=0  }  
 a_{k_1} (t) \cdot 
 a_{k_2}^* (t) \cdot 
 a_{k_3} (t) + F_l(t) \, ,
\label{eq:master_equation_haus}
\end{align}
where the slow amplitudes $a_l$ at the frequency $\omega_l = \omega_0 + l \, \Delta  \omega$, as imposed by the Fabry-P\'erot resonator,
are defined via the expansion of the electric field %in the cavity
\begin{align}
 E(z,t) = 
 \sum_l a_l (t) \, \eu^{-\iu \omega_l t} 
 \, \eu^{\iu q_l z} + \text{c.c.} \, .
\label{eq:elect_pulse}
\end{align}
In Eq. (\ref{eq:master_equation_haus}) the real parameter $G_l$ represents the difference between the gain and loss of the mode $l$ in a complete round-trip through the cavity, 
$D_l$ is the group velocity dispersion of the wave packet, 
$\Gamma$ is the nonlinear self-amplitude modulation coefficient associated to a saturable absorber 
%and, then, to the passive mode locking,
and $\Delta$ is the self-phase modulation coefficient (responsible of the Kerr lens effect).
The noise $F_l(t)$ is generally assumed Gaussian, white and uncorrelated:
\begin{align}
 & \langle F_{k_1}^* (t_1) \, F_{k_2} (t_2) \rangle = 2 T \, \delta_{k_1 k_2 } \, \delta (t_1-t_2) 
 \, , &
 & \langle F_{k_1} (t_1) \, F_{k_2} (t_2) \rangle = 0 \, ,
 \label{eq:uncorrelated_noise}
\end{align}
being $T$ the spectral power of the noise, associated, e.g., to spontaneous emission. 
% and it is related to the effective temperature.

To obtain an effective statistical approach the gain saturation %, that ensure the stability of the system,
must be explicitly considered.
Indeed, when $G_l$ and $\Gamma$ are nonzero,
the total optical intensity $\mathcal{E} \equiv \sum_k |a_k |^2$
is not a constant of motion for the evolution of Eq. (\ref{eq:master_equation_haus}).
In this case the laser is stable because the gain decreases as the optical intensity increases \cite{Chenetal94}.
For standard lasers this is usually modeled assuming that the gain in the master equation 
Eq. (\ref{eq:master_equation_haus}) is given by $G_l = G_0 /(1+\mathcal{E}/E_{\text{sat}})$,
being $E_{\text{sat}}$ the saturation power of the amplifier.
To study the equilibrium properties of the model, in SLD it is assumed a simpler model:
at any instant the gain is supposed to assume the value that keeps $\mathcal{E}$ exactly 
a constant of motion \cite{GorFis02}.
This corresponds to assuming the \emph{spherical constraint}
%In this way the system evolves over the hypersphere 
$\mathcal{E} \equiv E_0$
during the evolution of the slow amplitudes $a_l$.
The relation between the thermodynamics in the fixed-power ensemble and a variable-power ensemble
is similar to the one between the canonical and grand canonical ensembles 
in statistical mechanics \cite{GatGorFis04}.
%The constraint $\mathcal{E} \equiv E_0$ will induce a correlation of order $N^{-1}$ in the noise $F_l$ 
%but in the thermodynamic limit $N \gg 1$, in which we are interested, such correlation can be neglected and 
%the noise can be still considered white for all the practical purposes.

In the limit of small dispersion $D_l \ll G_l$ and $\Delta \ll \Gamma$,
the evolution becomes Hamiltonian with 
\begin{align}
 \mathcal{H}
 =&
 -  \sum_{k=1, \ldots N} G_{k}  |a_{k}|^2
 - \frac{\Gamma}{2}
 \sum_ { k_1 - k_2 + k_3 - k_4 = 0  }  
   a_{k_1} \,
   a_{k_2}^* \,
   a_{k_3} \,
   a_{k_4}^* \, , 
\label{eq:PML_hamiltonian}
\end{align}
where, in the fixed-power ensemble, $G_l$ is constant and the condition $\mathcal{E} = E_0 = \epsilon N$ is preserved.
This model can be, then, directly investigated with standard methods of statistical mechanics
for a system at the effective temperature $T_{\text{ph}} = T / \epsilon^2 = \mathcal{P}^{-2}$,
%$T = \Gamma \epsilon^2 / \mathcal{P}^2$,
being $\mathcal{P}$ the pumping rate of the source.
In particular, the mean field theory (MFT) of the model is exact when the 
Frequency Matching Condition  (FMC)
$ k_1 - k_2 + k_3 - k_4 = 0 $
on the nonlinear term is neglected.
In this case the above mentioned first order transition is obtained between 
a continuous and a pulsed regime.

\begin{figure}
\begin{center}
\subfigure[ \label{fig:signal_a}
Signal of Eq. (\ref{eq:elect_pulse}) (top) and corresponding configurations plotted as Phase vs Frequency (bottom) 
for a laser system Eq. (\ref{eq:PML_hamiltonian}) with $500$ modes just 
above the critical pumping ($\mathcal{P}=1.692$, $\mathcal{P}_c=1.565(8)$).
]{
\resizebox*{7.2cm}{!}{\includegraphics{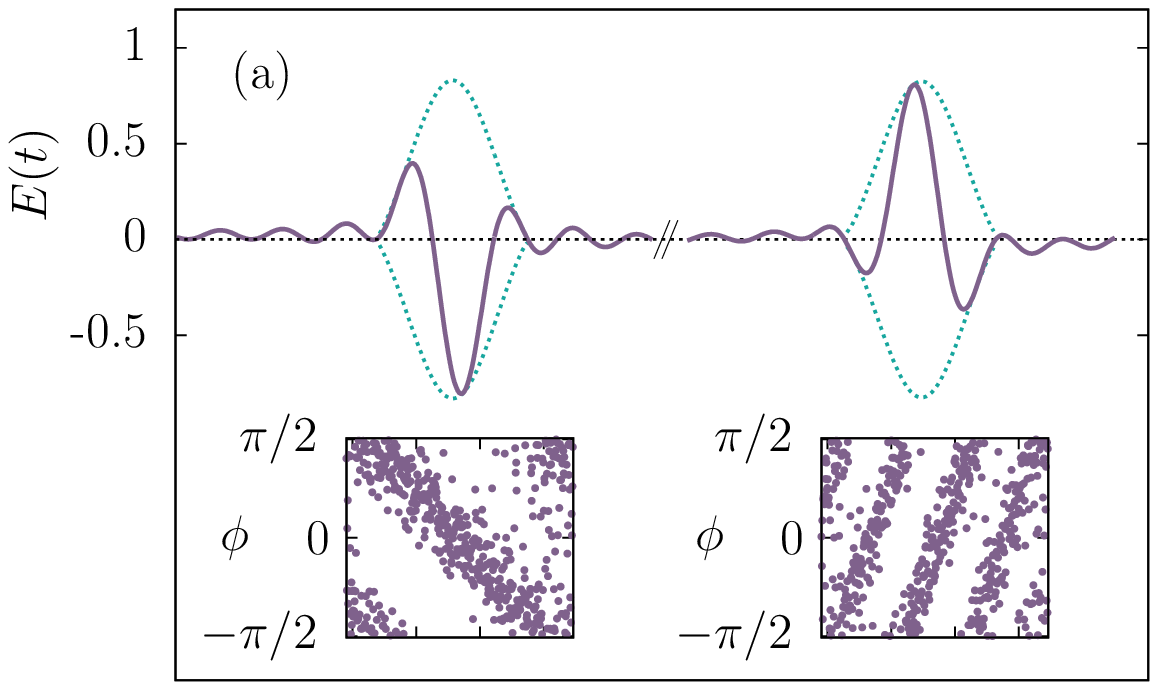}}}\hspace{30pt}
\subfigure[ \label{fig:signal_b}
Intensity spectra for three pumping for a laser system Eq. (\ref{eq:PML_hamiltonian}) 
with $150$ modes ($\mathcal{P}_c = 1.60(2)$ here).
Below the threshold the spectrum is shown to follow the gain curve $G(\omega_k) = G_k$ (black solid line).
]{
\resizebox*{7.2cm}{!}{\includegraphics{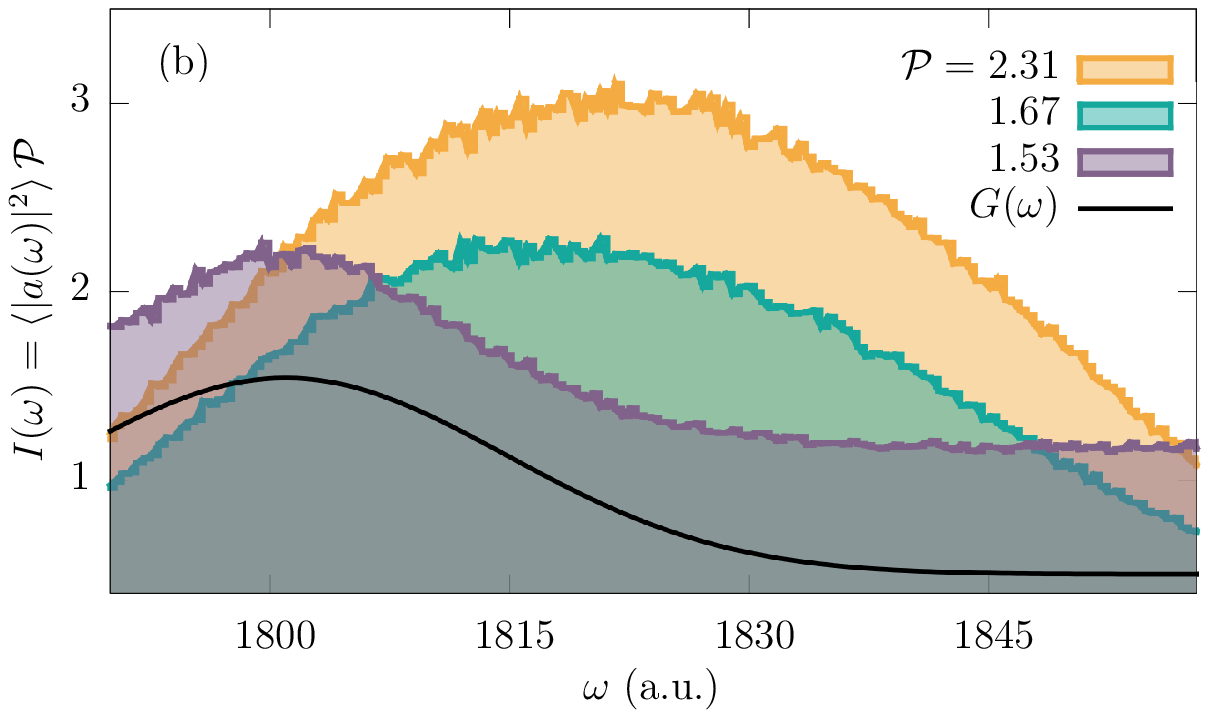}}}
\end{center}
\end{figure}

The role of the FMC in mode locked laser systems
has been the object of recent works 
within the statistical approach \cite{AntIbaLeu14b,MarLeu15,AntIbaLeu15}.
It is observed that, even in a laser system in the presence of FMC, a first order transition separates a continuous wave regime 
at low pumping and a pulsed regime at high pumping,
the critical pumping value being compatible in the thermodynamic limit with the MFT one. 
The difference between the MFT and FMC solutions is in the {\em nature} of the pulsed regime.
In MFT, the modes are all equivalent and they 
are trivially phase-locked with the same phase $\phi_j = \phi_0$ for $j=1 , \ldots N $ at high pumping.
Alternatively, when the FMC is included the pulsed regime is not trivially phase locked and
the laser mode phases result to behave as $\phi_j \simeq \phi_0 + \phi' \, \omega$, 
with a nontrivial, frequency independent, slope $\phi'$ (see insets in Fig. \ref{fig:signal_a}).
The slope $\phi'$ changes in the time evolution with a distribution determined by the interaction network among the modes 
and with a lifetime that increases with the pumping rate \cite{AntIbaLeu14b,AntIbaLeu15}.
Consequently, all the two mode correlators are zero in the high pumping regime in FMC systems \cite{AntIbaLeu15}.
The electromagnetic pulse associated with such configurations, cf. Eq. (\ref{eq:elect_pulse}),
%$E(t) = \sum_j |a_j| \exp \left[ \iu \left( \omega t + \phi_j \right) \right]$,
is, accordingly, such that the phase delay between the carrier and the envelope of the signal
changes at each shot, cf. Fig. \ref{fig:signal_a}.

The FMC turns out to play an essential role % not only for the phase locking %of the modes
also for the intensity spectra: 
for low pumping the spectrum simply follows the gain curve $G_k$ of Eq. (\ref{eq:PML_hamiltonian}), but,
as the pumping exceeds the mode locking threshold, the spectrum is determined mainly by the interaction network.
%It is, then, flat for low finesse systems while, for high finesse, 
In this case, then, it is observed to become narrow around the central frequencies of the bandwidth, cf. Fig. \ref{fig:signal_b}, as a direct consequence of the inhomogeneous connectivity induced by the FMC.
This furnishes a simple theoretical mechanism to explain the gain narrowing at the mode locking transition.

The transition from a continuous regime to a pulsed regime is observed also in models with fixed intensities \cite{MarLeu15,AntIbaLeu15}: 
the fluctuations of the amplitudes is neglected with respect to the one of the phases and the resulting Hamiltonian \eqref{eq:PML_hamiltonian}, 
obtained rescaling the interaction coefficients with the averaged amplitudes, 
can be mapped to the problem of non-linearly interacting $XY$ spins, $\sigma \equiv (\cos{\phi},\sin{\phi})$.
 %has been studied through Monte-Carlo simulations\cite{AntIbaLeu15}
%As a first step we    
 %Once the equilibrium configurations of the phases are known, it is then possible to reconstruct the electromagnetic field. 
% In \cite{AntIbaLeu15}, the dynamics of the phases is studied through Monte-Carlo simulations with system sizes up 
% to $10^2$ modes. Different degree of dilution, i.e., different values for connectivity per variable nodes are considered starting from the 
 %from here 
% The low temperature regime
% with correlated phases and absence of total magnetization 
% is observed as well indicating that the ``phase wave'', $\phi_j \simeq \phi_0 + \phi' \, \omega$, is completely
% determined by the dynamics of the phases.\\
 In Ref. \cite{MarLeu15}, 
 the equilibrium configurations of the phases are obtained in the thermodynamic limit mapping the system 
 on random regular graphs: when the number of modes is very high, the 
 details of the graph become unimportant and 
 the solutions are found averaging over many realizations of the network.
 %with same global characteristic, e.g., with same number of links on average. 
 The equilibrium configurations of the system are then studied 
 considering an extensive number of non-linearly interacting quadruplets. 
 This limit is interesting for RLs in which the spatial
 distribution of the modes could be highly localized. 
 Instead to employ the Monte Carlo method, one can then  
 study these diluted network systems using the cavity method, 
 widely used for optimization problems with discrete variable nodes. 
 The cavity method allows
 to determine the equilibrium distributions of the variable nodes starting from marginal probability distributions:
 to each link connecting a variable node and a function node are associated two probability distributions, 
 also called messages, representing
 the marginal probability distribution of one variable when the other is not there; e.g.,
 $
 \eta_{i \rightarrow a}(\phi)
 $
 indicates the message sent from variable node $i$ to function node $a$.
  The idea is that, since the graph is highly diluted, when a link 
 is cut the other variable nodes participating in that function node become uncorrelated; the problem can then be solved 
 self-consistently:
 $$
 \eta_{i \rightarrow a} = \mathcal{F}\{\eta_{b \rightarrow i}|_{b \in \partial i \setminus a}\}
 $$
 where with  $b \in \partial i \setminus a$ we indicate all the neighbors of $i$ except $a$. 
 Irregularity in the graph are considered evaluating the distribution of the $\eta$s.
 Non-trivial equilibrium configurations are obtained for low values of the effective bath temperature: together with the 
ferromagnetic phase, in which all the modes oscillates synchronously and that breaks the $O(2)$ symmetry,  
 a phase wave is observed. 
 The electromagnetic field is also reconstructed from the $\langle \phi_j \rangle$ and 
 the different regimes are analyzed also  respect
 to the starting frequency distributions. An electromagnetic pulse is obtained for a frequency comb distribution.
%For high pump values an impulsed regime is observed,   
%The phenomenon is observed also in models with fixed intensities 
%(XY and $p$-clock models) that include FMC \cite{MarLeu15,AntIbaLeu15} 

%%%%%%%%%%%%%%%%%%%%%%%%%%%%%%%%%%%%%%%%%%%%%%%%
\section{Multimode Laser Theory for Open and Disordered Lasers}
\label{sec:RL_optics}

After the pioneer works of Letokhov \cite{Letokhov68}, 
there was a renewed interest starting from the 90's  in light amplification in disordered systems both for
fundamental and applied research.
% 
%The study of light propagation and amplification in disordered systems started back in the early days of laser physics. In the late 60's, Letokhov\cite{Letokhov68}  firstly foresaw that
%light scattering, if high enough, could lead to light amplification and lasing: instead of being detrimental for laser action, multiple scattering would increase
%the time photons spend in the gain medium leading, under some circumstances, to lasing. To this phenomenon it was given the name of ``random laser'':
%light amplification by stimulated emission with feedback mediated by random space fluctuations of the dielectric constant, $\epsilon(\mathbf{r})$.
 %\red{After the  pioneer works of Letokhov \cite{Letokhov68}, the fields of light propagation and amplification in disordered media has 
 %seen an increase of interests since the 90's, as documented by the number of publications.} 
 The main features that distinguish RLs with respect to conventional 
 lasers are random disorder in the refractive index and absence of a light trapping cavity: 
 the confinement is realized by the multiple-scattering caused by the disorder itself. 
 %In this case the predominant loss mechanism for all the photons
In the first experiments it was 
unclear whether even this cavityless systems could have well-defined electromagnetic modes showing narrow peaks in the spectrum. 
% Random lasers are then highly open systems and the usual approximations of low leakage from the ``cavity'' do not apply in this case: the
 %openness make the problem non-Hermitian and the standard     
Indeed, light intensity
was described by a diffusion equation, with phases and hence interference not playing any role. In these cases, the threshold for lasing is obtained
 when the amplification length becomes smaller than the path length a photon travels before escaping from the gain medium. 
 Instead of 
 high-$Q$ resonances characterized by discrete peaks in the spectrum, 
 as in more conventional lasers, there appear low-$Q$ resonances spectrally overlapping: the spectrum is 
 continuous, it does not contain any discrete components at selected resonance frequencies and it is centered at the atomic transition, which is the only 
 remaining resonant element of the system. The feedback in these systems is indicated as intensity or energy feedback, it is incoherent and
  non-resonant (for a review on non-resonant feedback random lasers see Ref. \cite{Cao03}).
  
  In later experiments \cite{Cao05}, focusing higher intense pumps in smaller systems, it was possible to actually observe discrete lasing peaks in the intensity spectrum.
 These peaks were not necessarily centered around the atomic gain line and exhibited a Poisson photon count distribution, typical of coherent lasing light.
%  This spectrum is characteristic of conventional multi-modes lasers.
  Moreover, the position of the peaks
  was observed to change when a different part of the sample was illuminated.
  % and in same cases also shot-to-shot fluctuations were observed.
 % even when the incoming pump was illuminating the same spot.
   It was suggested that this transition in the  spectrum was due to a transition from a nonresonant to a resonant feedback: even though RLs are characterized by 
   the absence of a well-defined cavity yielding strong leakages, electromagnetic modes have enough time to form and lasing 
   involves coherent phase-sensitive feedback.

   After these new results, it has been accepted that RLs are multimode systems and research in this field has been focusing on the development of a multimode laser theory, 
   which, starting from the lines of a standard semiclassical multimode laser theory \cite{Lamb78}, has to incorporate specific
   features of these systems: strong radiative leakage, modes with a broad distributions of lifetimes and irregular spatial dependence of the refractive index;
    for three dimensional (3D) systems this implies also the absence of well-defined optical cavity axes.
    
    In the rest of this section we will briefly reference several alternative ways to introduce a multimode descriptions of strongly open systems. Starting 
    from a quantum treatment, we will then 
    analyze in more details the approach of Hackenbroich and co-workers \cite{VivHac04,Hackenbroich05} obtaining, under some hypothesis, a Hamiltonian of the form of Eq. \eqref{eq:PML_hamiltonian}.
    
\subsubsection*{Cold cavity modes in highly open systems}
    
    The problem of a modal description for open systems is of interest not only for RL physics and
    there have been introduced many descriptions for this purpose. 
    The difficulties of the problem increase due to the fact that
    leakages make the system non-Hermitian and the usual technique of developing a basis of orthogonal eigenvectors with real
     eigenvalues does not apply.   
    Concerning laser physics, Fox and Li \cite{Fox61} firstly studied the effect of losses for the electromagnetic modes in a cavity. In particular,
     they showed how the low-order modes, i.e. the modes that have lowest threshold, could be determined by a numerical iterative method: they 
     assumed an arbitrary field distribution at the first mirror and proceed to compute its distribution after one passage in the cavity; this distribution is 
     then used to compute the field at the second mirror and so on. 
The modes are then determined by the steady state distributions solving the self-consistent equations:
\begin{align}
\label{eq:fox_li}
  \int_{A} K\left(\mathbf{r}_\bot,\mathbf{r'}_\bot\right) \psi_n(\mathbf{r'}_\bot) d \mathbf{r'}_{\bot} = \lambda_n \psi_n(\mathbf{r}_{\bot}) 
\end{align}
   where $A$ is the mirror surface, $K$ is the kernel determined through the Huygens-Fresnel formulation and $\bot$ indicate the direction transverse to 
   the optical axis; $\lambda_n$ is a complex constant whose module and phase determine the attenuation and phase shift per transit for the mode.
   The application of Fox-Li modes for random lasing physics is rather limited for several reasons. First, in the 3D case,
    RLs lack of a well-defined optical axis and, more in general, the convergence of the self-consistence equations, Eq. \eqref{eq:fox_li},
     becomes slow and less effective
    when there are a lot of modes, which is the typical case for RLs.

    In general, the approach adopted in laser physics to predict the lasing modes is to start from the cold cavity modes, 
    which are solutions of the Maxwell equations with proper boundary conditions. %, as the Fox-Li modes.
    % The problem consists, then, to determine the correct boundary conditions. 
    In order to describe leakages, there have been introduced the so-called \emph{quasibound states} \cite{TurStoCol06}:
 solutions of the  Maxwell equations without gain and with boundary 
    conditions containing only outgoing waves at infinity. The Fox-Li method previously described is a technique to find the quasibound states of the system.  
    % which are modes of the system 
    %only defined inside the cavity.
     %quasibound states are solutions of Maxwell equations with boundary 
    %conditions containing only outgoing waves at infinity;  
    The quasibound states are, though, only defined inside the cavity: 
    they are associated
     to complex eigenvectors, see Eq. \eqref{eq:fox_li}, and, if extended outside the cavity, diverge at infinity, since they satisfy Maxwell 
     equations with only outgoing wave. 
     For this reason, quasibound modes cannot be used for a 
    quantization procedure in the entire space.

    The standard procedure to go beyond the cold cavity modes is to use the semiclassical laser theory, including the gain medium described
    as a two levels atom. The atoms are pumped to the upper level and then can decay to the lower level emitting a photon in the mode field; other atomic levels are in general 
    introduced through a finite lifetime of the two laser levels. The pumped medium is coupled to the field satisfying the Maxwell equations. The equations 
    obtained are the  
    Maxwell-Block equations (see, e.g., Ref. \cite{Lamb78} for the case of a cavity with a well-defined optical axis). One usually finds, for high-$Q$ cavities, 
    that the mode with highest $Q$ lases first with a frequency shifted toward the atomic transition. 
    The spatial distribution of the mode is not supposed
    to be altered much by the gain medium, i.e., it is basically the same distribution of the cold cavity mode. 
    The difficulties of the Maxwell-Block equations
    comes from the presence of multimodes: in this case nonlinearities can develop in the system and the equations cannot be solved exactly.
    Furthermore, even disregarding nonlinearities,
    a legitimated hypothesis only below the threshold, 
    the interactions are nontrivial in presence of many modes
    since they will depend on the (complex) modes spatial overlap.
%    this interaction is a direct consequence of the fact that the modes compete for the same gain medium.

    In respect to high-$Q$ cavities, 
    for the case of highly open systems with a nonuniform refractive index, 
    the lasing modes can be quite different from the cold cavity modes. One approach that has been used in this case is to determine
    the spatial distributions of the lasing modes as well as their frequencies solving Maxwell-Block equations self-consistently.
    %The idea
    %is that in this case the spatial distributions of the lasing modes as well as their frequencies 
    %should be determined solving Maxwell-Block equations self-consistently.  
    In Ref. \cite{TurStoCol06} the authors apply this approach disregarding nonlinearities caused by possible beat frequencies, 
    i.e.  neglecting population pulsation.
     To analytically obtain the solution, they use the Green function written in terms of a new set of eigenfunctions, the 
    \emph{constant flux states}. 
    Constant flux states are similar to the quasibound states, having a complex wave vector inside the cavity, 
    but differ from the latter because they have a real wave vector outside the cavity 
    not carrying a diverging flux at infinity 
    %(from which the name ``constant flux''). 
    (whence the name ``constant flux''). 
    Constant flux states solve different differential equations inside and 
    outside the cavity; continuity determines boundary conditions.
    Considering the case of a one-dimensional slab, 
    the set of the constant flux states $\{\pmb{\phi}_m(x,\omega)\}$ is defined to satisfy
\begin{align}
    - \nabla^2 \pmb{\phi}_m(x,\omega) = \mathbf{n}^2(x,\omega) \left(k_m + k_a\right)^2\pmb{\phi}_m(x,\omega)
\end{align}
    inside the cavity, while 
\begin{align}
     - \nabla^2 \pmb{\phi}_m(x,\omega) = \left(k + k_a\right)^2\pmb{\phi}_m(x,\omega)
\end{align}
    outside the cavity with outgoing wave boundary conditions.
    Here $\mathbf{n}(x,\omega)$ is the refraction index of the material and $k_a = E_a \, (\hbar c )^{-1}$, being $E_a$ 
    the atomic transition energy.
    Constant flux states are not orthogonal or complete. 
    Due to the non-Hermitian nature of the problem, 
    to obtain a complete set, 
    it is also necessary to introduce
    the functions $\{\overline{\pmb{\phi}}_m\}$, which are the adjoint of the constant flux states.
    These correspond to the solution 
    for free propagation outside the cavity with incoming waves only. 
     Constant flux states with their adjoint functions form a 
     complete set that can be used for a quantization in the whole
     space: the amplitudes become operators, and two set of operators are used inside and outside the cavity. 
     However, constant flux states are not very convenient 
     as a basis for a quantization procedure since the 
     internal and the external 
     operators do not commute; non-commuting relations  
     introduce coupling between the inside and the outside operators. For this reason, in the rest of this section, 
     we will introduce the system-and-bath-approach as a starting point for 
     a quantization procedure.
     
     \subsubsection*{The system-and-bath approach}
     
     The idea of the quantization technique based on the system-and-bath-approach is to separate the space in 
     two subsystems introducing separate operators 
     and eigenmodes. One subsystem is the open cavity, the resonator, containing the gain medium; 
     the open surrounding space  
     constitutes the bath. The bath and the resonator interact at the boundary. 
     This technique is adequate to study  
     the field degrees of freedom in the resonator: 
     at the end of the quantization procedure the degrees of freedom of the bath
     can be averaged out, obtaining then information
      % on the dynamics of the resonator.
       %This procedure is more suitable in this case 
      %since other methods, such as the modes-of-the-universe approach, do not provide explicit information
       about the field inside the resonator alone. % the bath introduces both losses and noise.\\
      
      The system-and-bath model is in general a standard description of losses 
      in quantum mechanical systems \cite{Lamb78,ScuZub01}
      % The 
      %bath, or reservoir, introduces both losses and noise.
      and it has been used for more than 40-years for quantum optical systems. 
      Some well-known,
      textbook examples,  
      are the derivations of the finite lifetimes of the excited levels of an atom and the damping of 
      a ``single'' oscillator mode. Less understood 
      is instead the damping in the case of multimode fields. In this case, the system is modulated by 
      a discrete set of independent harmonic oscillators, obtained from the
       quantization of the normal modes of the system. A continuum set of harmonic oscillators models 
       the outside reservoir. Coupling between the two gives rise to damping. 
    However, in this description, coupling constants only enter as phenomenological 
    parameters and it has been argued that it can be 
    applied only to systems with low coupling with the outside reservoir, as more conventional lasers.
Hackenbroich and coworkers have shown 
in a series of papers \cite{HacVivHaa02,VivHac03,HacVivHaa03} how to 
rigorously derive the system-and-bath Hamiltonian starting from Maxwell equations employing 
the Feshbach projector formalism \cite{Feshbach62},
a well-known technique in condensed matter and nuclear physics, 
hence providing a microscopic description of the theory. 
        In Ref. \cite{VivHac04}, they also derived 
        explicitly the expression for the coupling constants for some simple geometries. 
        We will briefly sketch their results.

        Let us consider the case of uniform 
        three-dimension linear dielectric medium, $\epsilon(\mathbf{r})$. %; the gain will be considered later. 
        The medium is surrounded by 
        free space, $\epsilon(r)=1$. In the absence of source, the eigenmodes $\mathbf{f}_m(\omega,\mathbf{r})$ 
        of the whole system satisfy:
        \begin{equation}\label{eq:modes_universe}
        \nabla \times \left[ \nabla \times \mathbf{f}_m(\omega,\mathbf{r}) \right] - \frac{\epsilon( \mathbf{r} )\omega^2}{c^2}%
         \mathbf{f}_m(\omega,\mathbf{r}) = \mbf{0}
        \end{equation}   
       the $\mathbf{f}_m(\omega,\mathbf{r})$ are defined in the whole space, $\omega$ is a continuous 
       parameter while the discrete index $m$
       indicates the boundary conditions depending on the problem at hand.
        For example, for a dielectric coupled to free space $m$ indicates the
       possible polarizations. In order to transform Eq. \eqref{eq:modes_universe} into an eigenvalues problem, it is convenient to introduce 
       the set of functions, $\mbf{\phi}_m(\omega,\mbf{r}) = \sqrt{\epsilon(\mbf{r})} \mbf{f}_m(\omega,\mbf{r})$; the 
       $\mbf{\phi}_m(\omega,\mbf{r})$ are the solution of the equation
       \begin{equation}\label{eq:modes_universe_t}
       L \mbf{\phi}(\omega,\mbf{r}) \equiv \frac{1}{\sqrt{\epsilon(\mbf{r})}} \nabla \times \left[ \nabla \times \frac{\mbf{\phi}(\omega,\mbf{r})%
       }{\sqrt{\epsilon(\mbf{r})}}\right] = \left(\frac{\omega}{c}\right)^2 \mbf{\phi}(\omega,\mbf{r}) \, .
       \end{equation}
       The functions $\mbf{\phi}_m(\omega,\mbf{r})$ form a complete set in the subspace of $L^2$ 
       functions defined by the condition
       $
       \nabla \cdot \left[ \sqrt{\epsilon(\mbf{r})}\mbf{\phi}_m(\omega,\mbf{r})\right]=0 
       $.
       The vector potential $\mathbf{A}$ can be expressed as a superposition of the  $\mbf{\phi}_m(\omega,\mbf{r})$
       functions with proper coefficients
       \begin{equation}\label{eq:vec_a}
      \mbf{A}(\mbf{r},t)  =  c \int d \omega \sum_m q_m(\omega,t) \sqrt{\epsilon(\mbf{r})}\mathbf{\phi}_m(\omega,\mbf{r}) \, ;
       \end{equation}
       same for 
       the canonical momentum, 
       $\mbf{\Pi}(\mbf{r}) =  \epsilon(\mbf{r}) \dot{\mbf{A}}(\mbf{r})/c^2$,
       with coefficients $p_m(\omega,t)$. Quantization is now obtained promoting 
       $q(\omega)$ and $p(\omega)$ as operators %; then, the Heisenberg equations of motions lead to Maxwell equations imposing the 
       with proper commutation relations \cite{VivHac03}.

       The quantization of the electromagnetic modes  starting from the eigenfunctions of the Helmholtz equations, Eq. \eqref{eq:modes_universe}, is 
       known as the \emph{mode-of-the-universe approach}. The resulting Hamiltonian is a sum of harmonic oscillators. However,
        as discussed before, we would like to have explicit information on the modes inside the resonators.
%        : we will then apply the system-and-bath approach. 
       %\mbf{\Pi}(\mbf{r},t) & = & \frac{1}{c} \int d \omega \sum_m q_m(\omega,t) \sqrt{\epsilon(\mbf{r})}\mathbf{\phi}_m(\omega,\mbf{r})
       %\textit{for Fabrizio: un piccolo passaggio formale che non capisco qui, come si calcolano i coefficienti $q_m$?; cio? capisco che poi diventano degli 
       %operatori, ma anche il loro valor medio dal punto di vista formale come si calcola?}\\
       In the system-and-bath approach, we consider then two subsystems: 
       one containing the resonator the other the open space outside.
       This can be implemented defining the two Feshbach projection operators \cite{Feshbach62}
       \begin{align}
       \mathcal{Q} & =  \int_{ r \in V} d \mbf{r} | \mbf{r} \rangle \langle \mbf{r}| \, , &
       \label{eq:q}
       \mathcal{P} & =  \int_{r \notin V} d \mbf{r} | \mbf{r} \rangle \langle \mbf{r}|  
       \end{align}
       representing the projection into the resonator and in the outside space respectively; 
       $V$ indicates the space occupied by the resonator.
       The operators $\mathcal{Q}$ and $\mathcal{P}$ satisfy: 
        $\mathcal{Q}=\mathcal{Q}^{\dagger}$,$\mathcal{Q}=\mathcal{Q}^2$, same for $\mathcal{P}$, 
        and are complete and orthogonal
       $\mathcal{Q}+\mathcal{P}=\mathcal{I}$, $\mathcal{Q}\mathcal{P}=\mathcal{P}\mathcal{Q}=0$. 
       The functions $\mbf{\phi} (\mbf{r})$ can now be decomposed
       through $\mathcal{P}$ and $\mathcal{Q}$ 
       \begin{equation}\label{eq:decomposition}
       \mbf{\phi}(\mbf{r}) = \mathcal{Q} \mbf{\phi}(\mbf{r}) + \mathcal{P}\mbf{\phi}(\mbf{r}) 
       = \mbf{\mu}(\mbf{r})  + \mbf{\nu}(\mbf{r}) 
       %\chi_{-}(\mbf{r})\mu(\mbf{r}) +  \chi_{+}(\mbf{r})\nu(\mbf{r})
       \end{equation}
       and Eq. \eqref{eq:modes_universe_t} can be expressed, then, as %using Eq. \eqref{eq:decomposition}:
       \begin{equation}\label{eq:mu_nu}
       \begin{pmatrix}
       L_{\mathcal{Q}\mathcal{Q}} & L_{\mathcal{Q}\mathcal{P}} \\
       L_{\mathcal{P}\mathcal{Q}} & L_{\mathcal{P}\mathcal{P}}
       \end{pmatrix}
       \begin{pmatrix}
       \mbf{\mu}(\mbf{r}) \\
       \mbf{\nu}(\mbf{r})
       \end{pmatrix}
        = \left(\frac{\omega}{c}\right)^2 \begin{pmatrix}
       \mbf{\mu}(\mbf{r}) \\
       \mbf{\nu}(\mbf{r})
       \end{pmatrix}
       \end{equation}
      with
     $ L_{\mathcal{Q}\mathcal{Q}}  \equiv  \mathcal{Q} L \mathcal{Q} $, 
     $ L_{\mathcal{Q}\mathcal{P}}  \equiv  \mathcal{Q} L \mathcal{P} $ and so on.
%      \end{eqnarray*}
     The operators $ L_{\mathcal{Q}\mathcal{Q}}$ and  $L_{\mathcal{P}\mathcal{P}}$ are self-adjoint operators in the Hilbert space of the resonator
      and outside channel functions, respectively. 
      Let us indicate the set of functions satisfying the eigenvalues problem,  
       with proper boundary conditions assuring continuity at the boundary, see \cite{VivHac03},
        with $\mu_{\lambda}$ for $ L_{\mathcal{Q}\mathcal{Q}}$,  and $\nu_m(\omega)$ for $ L_{\mathcal{P}\mathcal{P}}$. 
        The eigenfunctions of the resonator form a discrete set, which is complete and orthonormal within the resonator region. The eigenfunctions
        of the outside channel region from a continuum indexed by the frequency, $\omega$; the index $m$ specifies the asymptotic boundary
        condition. The functions $\nu_m(\omega)$, as the $\mu_{\lambda}$s, form a basis for the channel region.

        After the decomposition in the resonators and channel modes, 
        the quantization procedure can be carried out as usual: 
        the coefficients multiplying $\mu_{\lambda}$ and $\nu_m(\omega)$ in the decomposition for $\mbf{A}$ and 
        $\mbf{\Pi}$ 
        are promoted to operators with proper commutation relations 
        in order to satisfy 
        canonical commutation relations for $\mbf{A}$ and $\mbf{\Pi}$. 
        The system-and-bath Hamiltonian has, eventually, the form
        \begin{align}
        \nonumber
        \mathcal{H} = & \sum_{\lambda} \hbar \omega_\lambda  \, \alpha_{\lambda}^{\dagger} \,  \alpha_{\lambda} + %
        \sum_{m} \int d \omega \, \hbar \omega \,
        \beta_{m}^{\dagger}(\omega) \, \beta_{m}(\omega) + \\
        & \hbar \sum_{\lambda} \sum_m \int d \omega \left[ W_{\lambda m}(\omega) %
         \, \alpha_{\lambda}^{\dagger} \, \beta_m(\omega) + \mbox{h.c.}\right] \, .
        \label{eq:hamiltonian_first}
        \end{align}
        Here, the operators $\alpha_{\lambda}^{\dagger}$ and $\alpha_{\lambda}$ are the bosonic creation and annihilation operators, respectively,
         associated with the
        resonator mode $\lambda$; they satisfy $[\alpha_{\lambda}^{\dagger},\alpha_{\lambda'}^{\dagger}]=[\alpha_{\lambda},\alpha_{\lambda'}]=0$,
        $[\alpha_{\lambda},\alpha_{\lambda'}^{\dagger}]=\delta_{\lambda,\lambda'}$.
         Same for the operators $\beta^{\dagger}_m(\omega)$ and
        $\beta_m(\omega)$. The matrix $W_{\lambda m}(\omega)$ expresses the coupling between the inside and outside functions; 
        explicit expressions can be found in \cite{VivHac04}. 
        Note also that in writing Eq. \eqref{eq:hamiltonian_first} the rotating wave 
        approximations is assumed: we have kept only resonant terms $(\alpha^{\dagger}\beta,\beta^{\dagger}\alpha)$
         and disregarded nonresonant terms
        $(\alpha^{\dagger}\beta^{\dagger},\beta \alpha)$, which can become important when the widths of the modes are comparable to their frequencies.
Eventually, since we are interested in the dynamics of the resonator modes, we can obtain quantum Langevin
        equations for the internal modes only: the coupling with the external fields gives rise to damping and noise. 
        In the Heisenberg representation,
        the dynamical equations for the operators $\alpha_{\lambda}$ are:
        \begin{equation}\label{eq:langevin_first}
        \dot{\alpha}_{\lambda}(t) = -\iu \omega_\lambda \alpha_{\lambda}(t) 
        - \pi \sum_{\lambda'} \left[ W W^{\dagger} \right]_{\lambda \lambda'} \alpha_{\lambda'}(t)%
        + F_{\lambda}(t)
        \end{equation}
        where $F_{\lambda}(t)$, depending only on the $\beta_m(t)$ operators,  gives the noise term associated to the bath
        \begin{align}
        F_{\lambda}(t) = - \iu \int d \omega \eu^{-i\iu \omega(t-t_0)} \sum_{m}W_{\lambda m}\beta_{m}(\omega,t_0)
        \end{align}
and we are assuming the Markov approximation: the coefficient $W_{\lambda m}(\omega)$ are supposed to be independent
        of frequency over a sufficiently large frequency band around the atomic frequency \cite{Fyodorov97}.

        The result of Eq. \eqref{eq:langevin_first} generalizes to the case of many modes the Langevin equation 
         obtained for the damping of one mode in a cavity, see, for example, \cite{ScuZub01}. 
         There are two main differences with the equation known in laser theory: the noise operators have also non-diagonal correlations, 
         $\langle F_{\lambda}^{\dagger}F_{\lambda'}\rangle \neq \delta_{\lambda \lambda'}$, and the mode operators $\alpha_{\lambda}$ are not 
         independent oscillators, as in standard laser theory, but are coupled 
         through the damping matrix $\left[W W^{\dagger}\right]_{\lambda \lambda'}$. 
         Standard theory is recovered in the limit of 
         small leakages. i.e., small coupling with the outside reservoir.

\subsubsection*{Langevin equations for mode in random lasers}
         
 Having summarized the main results known for the dumping of multimodes fields, 
         we are now interested in considering the optical resonator filled
         with an active medium to actually describe a RL. 
         The medium is usually represented by a large number of homogeneously broadened two-level atoms whose density is indicated
         with $\rho(\mbf{r})$. The atomic transition, $\omega_a$, 
         is supposed to be much larger than the broadening of the non-diagonal 
         elements of the atomic density-matrix, $\gamma_{\bot}$ \cite{Lamb78}.
         In the Heisenberg representation, the atom-field dynamics is described through:
         \begin{align}
         \dot{\alpha}_{\lambda} & =  - \iu \omega_{\lambda} \alpha_{\lambda} - \sum_{\mu} \gamma_{\lambda \mu} \alpha_{\mu} + %
         \int d \mbf{r} g^{\dagger}_{\lambda}(\mbf{r}) \sigma_{-}(\mbf{r}) + F_{\lambda} \label{eq:dyn_mode} \\
         \dot{\sigma}_{-}(\mbf{r}) & =  -(\gamma_{\bot} + \iu \omega_a) \sigma_{-}(\mbf{r}) + %
         2 \sum_{\lambda} g_{\lambda}(\mbf{r})\sigma_z(\mbf{r}) \alpha_{\lambda} + F_{-}(\mbf{r}) \label{eq:sigma_} \\
         \dot{\sigma}_{z}(\mbf{r}) & =  \gamma_{\parallel} \left(S \rho(\mbf{r}) - \sigma_z(\mbf{r})\right) - %
         \sum_{\lambda} \left(g^{\dagger}_{\lambda}(\mbf{r})\alpha^{\dagger}_\lambda\sigma_{-}(\mbf{r}) + \text{h.c.} \right) + %
         F_{z}(\mbf{r}) \label{eq:sigma_z} 
         \end{align}
          where $\sigma^{\dagger}=|e\rangle \langle g|$ and
           $\sigma_{-}=| g \rangle \langle e|$ are the atomic raising and 
          lowering operator, respectively. $|e\rangle$ and $|g\rangle$ represent the excited and ground states of the atom; 
          $\sigma_z=|e\rangle \langle e| - |g\rangle \langle g|$, is the inversion operator, $\gamma_{\parallel}$ its decay rate.
           The atom density is $\rho(\mbf{r})$, while $S$ indicates an external pump injecting atoms in the excited level.
           %In order to describe the coupling between the atoms and the fields we suppose valid the dipole approximation;
           In the dipole approximation, which we assume to be valid, the atom-field couplings are:
\begin{align}
           g_{\lambda}(\mbf{r}) = \frac{\omega_a p }{\sqrt{2 \hbar \epsilon_0 \omega_{\lambda}}}  \mu_{\lambda}(\mbf{r})
\label{eq:def_g}
\end{align}
           being $p$ the atomic dipole matrix element: $ p \propto  \langle e \mbf{r} g \rangle$. To shorten 
           the notation, we have replaced the coupling matrix of Eq. \eqref{eq:langevin_first} 
           with $\gamma_{\lambda \mu} \equiv \pi  \left[ W W^{\dagger} \right]_{\lambda \mu}$.

Semiclassical laser theory consists in neglecting noise terms and
replacing operators with their expectation values. We consider
laser media for which the characteristic times of atomic pump and loss
are much shorter than the lifetimes of the photons in the
resonator. In this
case the atomic variables can be removed obtaining nonlinear equations for the field alone,
equivalent to the master equation for standard lasers, Eq. (\ref{eq:master_equation_haus}).
%
%As outlined before, we are interested in the multimodes regime: the
%system is above lasing threshold, many modes develop in the system and
%the time dependence of the atomic operators cannot be neglected. 
%In this limit 
The standard procedure is to consider an expansion in mode
amplitudes, $\alpha_{\lambda}$: one starts neglecting the quadratic
term in Eq. \eqref{eq:sigma_z}, obtaining the zero-order
approximation, which, replaced in Eq. \eqref{eq:sigma_} gives the
first order approximation that replaced back in Eq. \eqref{eq:sigma_z}
gives the second order approximation and so on. 
We will limit to third
order terms; subsequent terms may become important far above threshold
and, from a statistical mechanical point
of view, they are not expected to change the universality class of the transition, see, e.g. \cite{CriLeu13}.
%
%
%Above threshold, several modes, labeled by index $k$, may perform laser action. 
Eventually, the Langevin equations for the mode amplitude take a simple form
in the basis of the \emph{slow amplitude modes}.
For these modes $\overline{a}_k$ we make the ansatz \cite{Hackenbroich05,AntCriLeu14c}:
\begin{align}\label{eq:slow}
    \alpha_{\lambda} =& \sum_k A_{\lambda k}\overline{a}_k(t) 
    \, , & \text{with } \qquad 
    \overline{a}_k(t) =& a_k(t) e^{-i \omega_k t}
\end{align}
%for yet unknown frequencies $\omega_k$. 
where the amplitudes
           $a_k(t)$ evolve on time scales much larger than the
           oscillation period $\omega_k^{-1}$.
           %, for this reason, these
           %solutions are known as the slow amplitude
           %modes \cite{AntCriLeu14c}. 
           %By definition, in the regime of
           %resonant feedback, a lasing mode is a slow amplitude mode
           %with positive amplitude, $\langle a_k(t) \rangle>0$. In
           %general, the number of lasing modes will increase with the
           %pump intensity.  Using the decomposition of
           %Eq. \eqref{eq:slow}, we can rewrite
           %Eqs. (\ref{eq:dyn_mode},\ref{eq:sigma_},\ref{eq:sigma_z})
           %obtaining dynamical equations
           %for the lasing mode amplitudes, see \cite{AntCriLeu14c}.\\
           These modes play the role of the normal modes in a Fabry-P\'erot resonator for a standard laser, 
           see Eq. \eqref{eq:elect_pulse}, 
           and their nontrivial structure is a result of the complex multiple-scattering 
           that traps the light in this case.
           In analogy with the standard case, cf. Eq. \eqref{eq:PML_hamiltonian},
the complex Langevin equations for the
           amplitude $a_k(t)$ can be written then as
           \begin{align}
\dot{a}_k(t) =  -\frac{\partial \mathcal{H}}{\partial a^*_k} 
           \end{align}
           with
           \begin{align}
           \mathcal{H} =  - \frac{1}{2} \sum_{\mbf{k}|\text{FMC}(\mbf{k})} g^{(2)}_{k_1 k_2}a_{k_1} a^*_{k_2}+%
           - \frac{1}{4!}\sum_{\mbf{k}|\text{FMC}(\mbf{k})} g^{(4)}_{k_1 k_2 k_3 k_4} a_{k_1}a^*_{k_2}a_{k_3}a^*_{k_4} 
           \label{eq:langevin_slow}
           \end{align}
           where the sums are restricted to terms that meet the FMC which,
            for generic $2n$ interacting modes, reads
\begin{align}
           \text{FMC}(\mbf{k}) = %
           \text{FMC}(\mbf{k}_1, \dots , \mbf{k}_{2n}): \quad
           |\omega_{k_1} - \omega_{k_2} + \dots + \omega_{k_{2n-1}} - \omega_{k_{2n}}| < \gamma
\label{eq:FMC}
\end{align}
           with $\gamma$ being the typical linewidth of the mode. 
           The linewidth can be determined from a full quantum
           theory, in the following it will be supposed as a
           parameter.  
           The couplings are $g^{(2)}_{k_1 k_2} = \tilde{\gamma}_{k_1 k_2 } - S\, G^{(2)}_{k_1 k_2}$           
           and $g^{(4)}_{k_1 k_2 k_3 k_4} = 2 S G_{k_1 k_2 k_3 k_4}^{(4)}$,
           being $\tilde{\gamma}_{k_1 k_2} = \sum_{\lambda \mu} A^{-1}_{\lambda k_1} \gamma_{\lambda \mu} A_{\mu k_2}$
           and
           \begin{align}
\label{eq:G2_coldcavity}
  G_{k_1 k_2}^{(2)} \propto &   \int d \mathbf{r} \, \rho(\mathbf{r}) 
  \, g_{k_1}^{L *} (\mathbf{r}) \, g_{k_2}^{R} (\mathbf{r})
 \, , &
 G_{k_1 k_2 k_3 k_4 }^{(4)} \propto &
 \int d \mathbf{r} \rho(\mathbf{r}) 
 g^{L *}_{k_1} (\mathbf{r})  g_{k_2}^R (\mathbf{r}) 
 g_{k_3}^{R *} (\mathbf{r})  g_{k_4}^R (\mathbf{r})  
\end{align} 
where where the  coefficients proportionality slightly depending
on the frequencies involved \cite{AntCriLeu14c},
$g_k^L = \sum_\mu \left( A^{-1} \right)^{*}_{\mu k} g_\mu$
and $g_k^R = \sum_\mu A_{\mu k} g_\mu$ (cf. Eq. \eqref{eq:def_g}).
Hence, 
we see the main modifications brought by random lasers:
the openness of the cavity
introduces the linear interaction $\tilde{\gamma}_{k_1 k_2}$, 
due to the coupling of all the modes to the same external bath,
while the disordered structure of the modes establishes
nontrival, positive and negative, interactions $G_{\mathbf{k}}^{(p)}$,
due to the coupling of spatial overlapping modes with the same gain medium.

           Up to now, in the semiclassical approach, we have
           completely disregarded the effect of noise. However, for a
           complete statistical mechanical description, noise terms
           must be taken into account providing an effecting
           bath-temperature describing random oscillations.  We have
           seen before that different sources of noise, $F_\lambda$,$F_z$ and $F_-$, 
           occur in Eqs. \eqref{eq:dyn_mode}-\eqref{eq:sigma_z}; and, for open cavities, the noise
           terms, $F_{\lambda}$, have non-diagonal correlations. The
           decomposition in the slow amplitude modes,
           Eq. \eqref{eq:slow}, affects also the noise
           terms \cite{AntCriLeu14c}; this decomposition is by no means
           unique \cite{EreSkiOrs11} and this freedom may be used in
           order to build a mode basis where the noises do not have
           non-diagonal correlations.
           In general this could be done at the cost to have a non-diagonal linear interactions $g_{k_1 k_2}^{(2)}$. 
           In the rest of this study, we
           will consider a situation in which such decomposition is possible
           and, then, that the noise is white and uncorrelated
           as in the standard laser case, see Eq. \eqref{eq:uncorrelated_noise}.
In this situation, the same theoretical framework adopted for SLD, cf. section \ref{sec:ordered},
can be preserved for general multimode systems:
in the limit of small dispersion the evolution becomes Hamiltonian
and, assuming the spherical constraint $\mathcal{E} = E_0$,
the gain can be considered stationary. 
This sets up the investigation of random lasers through a statistical mechanics 
study of the Hamiltonian Eq. \eqref{eq:langevin_slow}.

%%%%%%%%%%%%%%%%%%%%%%%%%%%%%%%%%%%%%%%%%%%%%%%%
\section{A glassy laser}
\label{sec:RL_statMec}

Among the most singular aspects of RLs is that, at least for some photonic systems composed by a large number 
of modes, a complex behavior in its temporal and spectral response is observed:
if there is no specific frequency that dominates the others, 
%the laser can have a different spectrum each time it is excited  \cite{39, 41?, 42?}.
the narrow emission spikes in the spectra can change frequency from one excitation pulse to another 
with emission spectra that appear different from \emph{shot-to-shot} \cite{Mujetal07,Demetal04,vanderMolen06}.
In these cases the scattering particles and all the other external conditions are kept perfectly constant,
so these differences are only due to the spontaneous emission from which the RL starts at each shot.
In these conditions it is observed that the intensity distribution is not Gaussian, but rather of the Levy type 
%\footnote{ 
%Levy distributions are characterized by a slowly decaying (power-law) tail,
%so they have infinite variance and describe the occurrence of rare but very large values.
%}.
\cite{Sharmaetal06,Leprietal07}.
This is true close to the lasing threshold, 
whereas far below and far above the threshold the statistics remains Gaussian.
The shot-to-shot fluctuations provide also a possible explanation for the absence of narrow spikes 
%reason that
in some  experimental studies:
%have missed the observations of narrow spikes: 
if several emission shots are averaged over,
with the laser being in the above-mentioned complex regime, the spikes can be averaged out \cite{Wiersma08}.
Such peculiar behavior demands a theoretical explanation. 
%In the following we will discuss that 
%it can be understood as a basic manifestation of a corrugated free energy landscape.

%In such puzzling theoretical and experimental situation surrounding 
%\red{concerning/regarding/ [oppure leverei tutto da qua fino alla virgola]} the RL phenomenon, 
A major benefit can be obtained 
by an innovative point of view based on a statistical mechanics method, analogous to the SLD for mode locking lasers described in section \ref{sec:ordered}.
In this case the theory can profit of the great progress 
made in the last decades in the theory of complex systems.
The peculiarity of this approach is that 
it does not try to predict the properties of a given complex system, 
%as \emph{``a system is complex if its behavior crucially depends on the details of the system''} \cite{parisiComplex2002},
since \emph{``its behavior crucially depends on the details of the system''} \cite{Parisi02},
but the probability distribution of those properties for systems belonging to a given class.
In this case, one is not interested in the exact values of the couplings in the 
Hamiltonian Eq. \eqref{eq:langevin_slow} in a specific setting but only in their general statistical properties.
In particular, we point out as in the following the first two moments of the coupling probability distributions
define alone the thermodynamic behavior of the system.

In particular, the \emph{replica method} \cite{MPVBook}
has been adopted to study photonic system based on the Hamiltonian Eq. \eqref{eq:langevin_slow}.
In this approach, to probe the multi-state thermodynamic phase space, 
one considers $n$ copies of the
system with exactly the same set of disordered couplings (\emph{replicas})
and evaluates the disorder averaged partition
function $\overline{Z^n_J}$ of the replicated system. 
A continuation to \emph{real} $n$ is then considered to evaluate the free energy $F$
\begin{align}
  F =
 - T \, \overline{\log Z_J}
 = \lim_{n\to 0} \frac{\overline{Z^n_J}-1}{n} \, .
\end{align}
As the size of the system grows sufficiently large,  
the sample-to-sample fluctuations of the free
energy die out so that it becomes independent of disorder,
i.e., it is \emph{self- averaging}. 
In the MFT one is finally able to get rid of ``spatial'' dependency (corresponding to the frequencies in the photonic case)  and 
express the free energy as a function only of 
the (complex, replica independent) \emph{parameter of global coherence} $m = \sqrt{2} \sum_k a_k / N$,
analogous to the magnetization for spin models,
and the (real) \emph{generalized overlap matrices}
\begin{align}
\label{eq:def_QR}
 & Q_{\sfa \sfb} = \frac{1}{\mathcal{E}} \, \sum_{l=1}^N 
 \Re \left[ a^{\sfa}_l \left( a^{\sfb}_l \right)^\star \right]
 \, , &
 & R_{\sfa \sfb} = \frac{1}{\mathcal{E}} \sum_{l=1}^N 
 \Re \left[ a^{\sfa}_l  a^{\sfb}_l  \right] \, .
\end{align}
For $N \to \infty$, following the Parisi ansatz, the physical value
of the matrices follows from the extremization of the free
energy functional in the space of the $\mathcal{R}$-step RSB matrices \cite{MPVBook}.

A crucial point is that the mean field theory is exact in realistic physical situations for RLs.
This happens if the probability distribution of the couplings is the same for all the mode couples $(k_1, \, k_2 )$ 
and all the mode tetrads $(k_1 , \, k_2 , \, k_3, \, k_4 )$. 
Such circumstance is, e.g., the one of a RL that presents
\emph{extended modes} with a \emph{narrow-bandwidth}.
The last condition assures that $|\omega_1 - \omega_2 + \omega_3 - \omega_4| < \delta \omega$,
being $\delta \omega$ the typical linewidth of the lasing modes,
so that the FMC is always trivially satisfied.
Such physical circumstance is not possible in
more traditional applications of the theory of complex system,
like spin glasses and structural glasses, where the real system is inevitably \emph{short range}.
Indeed the nature of the glass phase of short range spin glasses is a debated problem (see, e.g., Ref. \cite{Young97}):
in this case the phenomenology of the system is
strongly affected by activated processes, such as nucleation and barrier crossing, 
that are, instead, negligible in the mean field limit.
Nevertheless, the degree of localization in the different experimental realizations of RLs 
ranges of several orders of magnitude, from strongly localized to modes extended all over the cavity:
RLs may then provide an interesting
cross-over between the mean field and short range systems that might be crucial for the understanding of 
glass phase in finite dimensions.

The deep connections
between theory of complex systems and disordered photonics
has been put forward
in the works of Angelani et al. \cite{Angetal06,Angetal06b} where they first show,
using a standard approach for quenched disordered systems based on the replica method, 
that the competition for the available gain of a large number of random modes
%(as, in particular, may be the situation where shot-to-shot fluctuations are observed)
can lead to a behavior similar to that of a glass transition \cite{Gotze92,Fisher91,MPVBook}:
at the lasing transition the photon gas
presents an exponential number of metastable states (i.e., it has a \emph{nonzero complexity}) 
corresponding to different mode locking processes in a RL.
In this case the leading mechanism for the non-deterministic activation of the modes, 
underlying the shot-to-shot fluctuations, 
%in the complex coherent wave regime % in nonlinear and strongly disordered systems
is identified in frustration and the induced presence of an exponential number of metastable states
\footnote{
we stress, nevertheless, that glassiness is in principle an independent phenomenon from deterministic chaos,
i.e. the high sensitivity to initial conditions, see, e.g., Ref. \cite{CriFalVul96}.
}.
All the rich phenomenology of glasses,
like aging, memory effects and history dependent responses \cite{Cugliandolo02},
is freshly predicted for nonlinear photonic systems.
In this case the relevant experimental quantities would be, in general,
the whole correlation functions in time domain 
\begin{align}
 C(t, \omega) = 
 \langle \, a_l(\tau+t) \, a_l(\tau) \, \rangle_\tau =
  \langle A_l (\tau+t) \, A_l (\tau) \, \eu^{\iu \left( \phi_l (\tau+t) - \phi_l (\tau) \right) } \, \rangle_\tau \, ,
 \label{eq:corrPhases}
\end{align}
where the average is over the time origin $\tau$.
They note that in molecular glasses the typical microscopic time scales are $\sim 10^{-12}$ $s$
while for the photonic dynamics are $\sim 10^{-14}$ $s$ (see, e.g., Ref. \cite{Sieetal98}),
so one could earn orders of magnitude to study and test theories of glass transition \cite{Biroli05}.
However, at our knowledge, the measure of the phase correlations required by Eq. (\ref{eq:corrPhases}) 
has not been achievable so far, the main complication being the typical low intensity 
of the RL emission,  
usually not intense enough to successfully
use techniques based on second-harmonic generation.

The analysis has been completed in the subsequent works \cite{Leuetal09,Conti11}
obtaining the whole phase diagram for an arbitrary degree of disorder,
bridging the results of SLD for passive mode locking (cf. section \ref{sec:ordered})
to the aforementioned of a completely disordered amplifying medium.
In particular, the ``glassy'' RL transition is present also for a negative value of the mean nonlinear interaction. 
In the thermodynamic construction, negative values correspond to
the absence of a saturable absorber, cf. Eq. (\ref{eq:master_equation_haus}), 
and, then, this would imply that mode locking processes in RLs are
achieved even without a saturable absorber.
In Ref. \cite{MarLeu15_try},
the same model is also analyzed using the cavity method.
We stress that this technique could
allow to include also the FMC in the description
when used on suited graph (cf. section \ref{sec:ordered}). 
Work in this direction is in progress.

\subsubsection*{Results for the general Hamiltonian model}

All the above mentioned results for the statistical approach to RLs 
are obtained from the same mean-field scheme
discussed in section \ref{sec:RL_optics} and leading to Eq. \eqref{eq:langevin_slow},
but using two additional approximations:
the \emph{strong cavity limit}
and 
the \emph{quenched amplitude approximation}.
The first considers the situation where the leakages from the cavity are very small,
so that their effects on the linear coupling and on the noise are negligible;
they, hence, only contemplate the nonlinear term in Eq. \eqref{eq:langevin_slow}.
The latter assumes a time scale separation between the 
evolution of ``slow'' amplitudes and ``fast'' phases of the modes;
in this case, they consider an effective Hamiltonian for the phases alone
while the amplitudes are fixed and have, then, just the role to renormalize the random couplings:
$g_{k_1 k_2 k_3 k_4}'  \equiv  |a_{k_1}| \, |a_{k_2}| \, |a_{k_3}| \, |a_{k_4}| \cdot g_{k_1 k_2 k_3 k_4} $.

In section \ref{sec:ordered} we have discussed as how the mode locking in standard lasers
is achieved as a nontrival locking of the mode phases (the so-called ``phase waves'', cf. Fig. \ref{fig:signal_a}).
In this mechanism the amplitudes play a minor role (de facto the same effect
is observed also for the ``quenched amplitude'' version of the model, 
see the results for the XY model in Ref. \cite{AntIbaLeu15}).
Nevertheless, even in the ordered case, the amplitudes have a nontrival dynamics
that leads to experimental predictions \cite{AntIbaLeu15}, cf. also Fig. \ref{fig:signal_b}.
The influence of mode amplitude dynamics is even more relevant in the case of RLs, where data about phase correlations
are hard to obtain (and not available so far, to our knowledge).
Moreover, the amplitudes are known to have, at least in some experimental configuration, a nontrival behavior,
as hinted by the shot-to-shot fluctuations of the intensity spectra.

To include the role of the leakages and the evolution of the mode amplitudes, 
a series of recent works %from some of the authors of this paper 
\cite{Antetal14,AntCriLeu14c,AntCriLeu15}
have considered the application of the replica method to the general Hamiltonian Eq. \eqref{eq:langevin_slow}.
This includes the possible presence of a linear term and the employment of
both the phase and the amplitude of the modes as degrees of freedom of the problem.
The FMC Eq. \eqref{eq:FMC} is, instead, still neglected to obtain a model for which the mean field theory is exact.
The full inclusion of the FMC requires the use of different techniques, as discussed
for the mode locking laser in section \ref{sec:ordered}.

In the mean-field theory the couplings $g^{(2)}_{k_1 k_2}$ and $g^{(4)}_{k_1 k_2 k_3 k_4}$ are taken as independent identically distributed random variables.
%From a statistical point of view, 
%In mean-field the behavior of the system is only determined by the first two moments of the probability distribution of the interactions. 
Without loss of generality, the probability distributions are considered Gaussian with ($p=2,4$)
\begin{align}
\label{eq:disorder}
& \mathcal{P} \left( g^{(p)}_{i_1 \ldots i_p} \right) 
\propto
%= \frac{1}{\sqrt{2 \pi \sigma_{p}^2} } 
 \exp \left[ - \frac{\left( g^{(p)}_{i_1 \ldots i_p} - \tilde{J}_0^{(p)} \right)^2}{2 \sigma_{p}^2} \right] 
  &  \text{with}  \quad 
 \sigma_{p}^2  = \frac{ p! \, J^2_p}{2 N^{p-1}} \, ,  \quad
  \tilde{J}_0^{(p)} =  \frac{J_0^{(p)}}{ N^{p-1}}\, .
\end{align}
The scaling of $\sigma_{p}$ and $\tilde{J}_0^{(p)}$ with $N$ assures the extensivity of the Hamiltonian.
To have a direct interpretation in terms of photonics quantities it is also useful to define the photonic parameters
\begin{align}
 J_0^{(4)} =& \alpha_0 J_0 
 \, , &
 J_0^{(2)} =& (1-\alpha_0) J_0 
 \, , &
  R_J = & \frac{J}{J_0} \, ,
 \\
 J_4^2 =& \alpha^2 J^2
 \, , &
 J_2^2 =& (1-\alpha)^2 J^2 
  \, , &
 \mathcal{P} = & \epsilon \sqrt{\beta J_0} 
 \, .
\end{align}
The parameters $J_0$ and $J$ fix, respectively, the \emph{cumulative strength} of the ordered and disordered contribution to the Hamiltonian while $\alpha_0$ and $\alpha$ 
the \emph{strength of nonlinearity} in the ordered and disordered parts.
The parameter $R_J$ is the \emph{degree of disorder}
and $\mathcal{P}$ is the \emph{pumping rate}, equivalent to the one defined for the mode locking laser in section \ref{sec:ordered}.

\begin{figure}
\begin{center}
\subfigure[ \label{fig:RL1_a}
Phase diagram in $\mathcal{P}$ vs $\alpha$ for a high degree of disorder $R_J$ (solutions with $m=0$) with static 
(solid) and dynamic (dashed) lines.  
%  The static lines are the static RFOT line between PLW to RL(1RSB) 
%  and RL(1RSB) to RL(1-FRSB) phases
%  with the 1RSB, 1-FRSB and FRSB structure.
 The dynamic line is where the dynamics arrests typical of a RFOT occurs 
  and an exponential number of metastable states appear.
% The solid black lines marks the appearance of the 1-FRSB solutions.
%  The dashed black line is the dynamic line for the 1-FRSB to 1RSB.
  On the solid lines between IW, PLW and RL(FRSB) the transition is continuous.
]{
\resizebox*{7.cm}{!}{\includegraphics{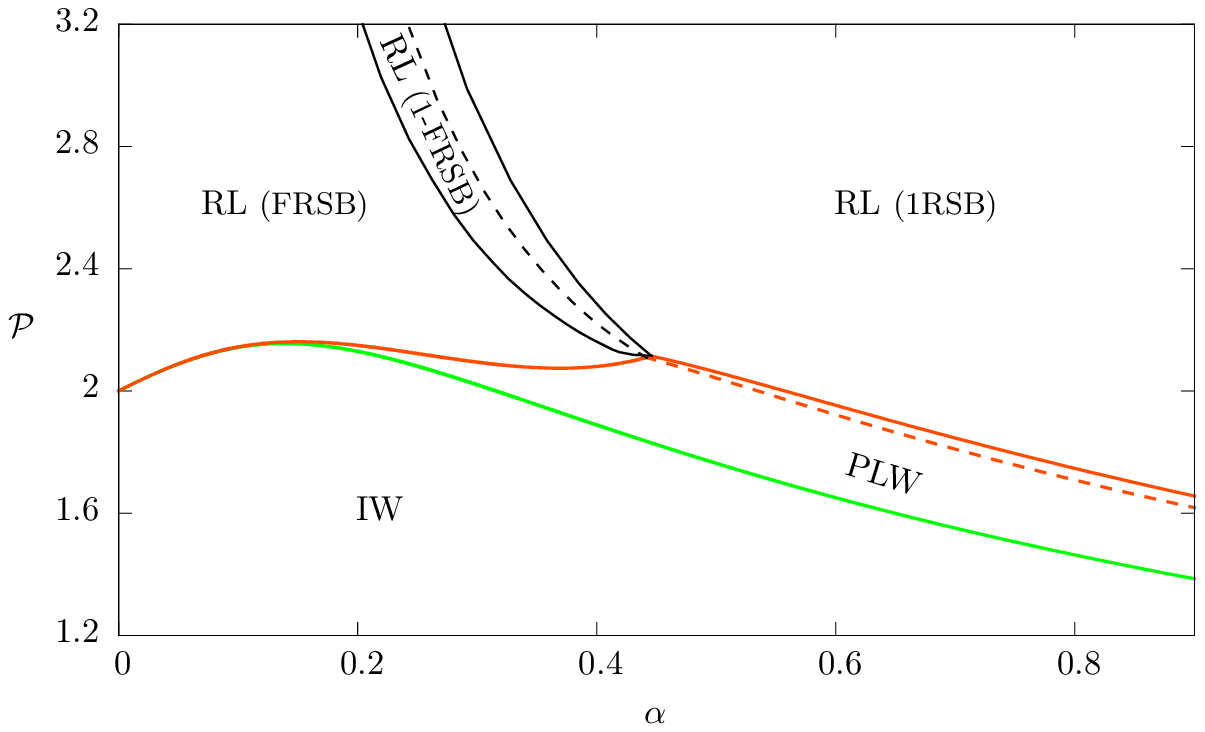}}}\hspace{40pt}
\subfigure[ \label{fig:RL1_b}
 Whole phase diagram in the photonics parameters for $\alpha = \alpha_0$.  The solid (dashed) red lines
  correspond to continuous (discontinuous) IW-ML transition.
  The blue surface is the RL-ML transition,
  the orange surface the PLW-RL transition
  and the green surface the IW-PLW transition.  
  The two
  black lines mark the intersection  between the
  orange-blue and green-red surfaces, respectively.
]{
\resizebox*{7.cm}{!}{\includegraphics{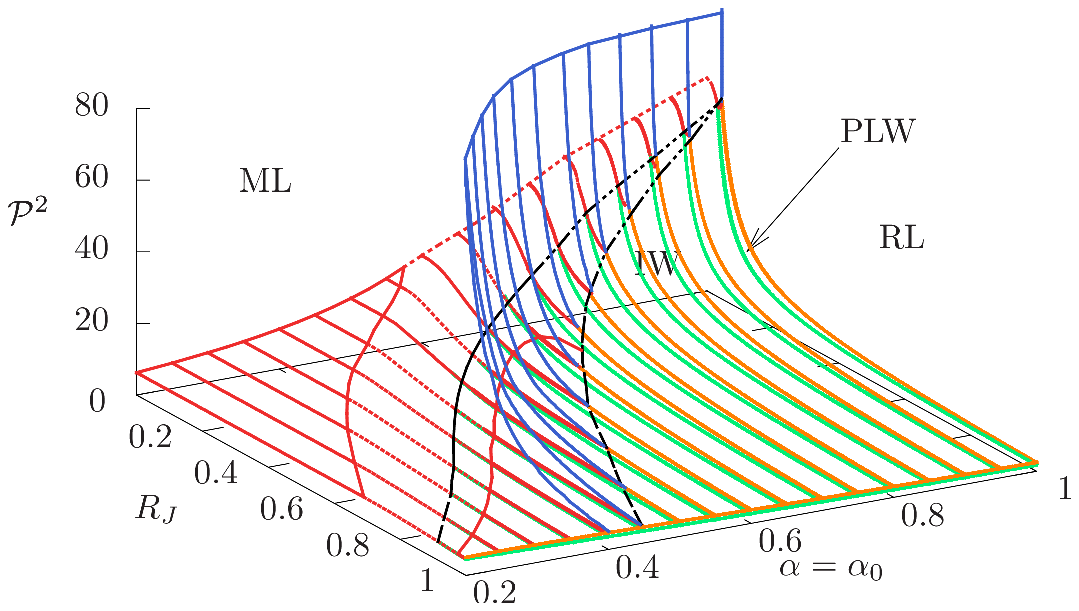}}}
\end{center}
\end{figure}

The whole phase diagram in terms of the order parameters $m$, $Q$ and $R$ 
(cf. Eq. \eqref{eq:def_QR}) is obtained in Refs. \cite{Antetal14,AntCriLeu14c} 
varying the previously mentioned photonics parameters.
A crucial point is that 
the simplest ansatz of assuming  $\Qp_{\sfa\sfb} = \Qp$ for all $\sfa \neq \sfb$,
 i.e., of assuming that all replicas are equivalent,
is not consistent in  the whole phase space.
%, specifically at high pumping.
Therefore, one  must allow for a  Replica Symmetry Breaking (RSB) to 
construct the solution, so that the elements of $Q_{\sfa \sfb}$ and $R_{\sfa \sfb}$ 
can take  different values and the order parameter becomes their probability distribution.
In this situation, then, identical copies of the system show different
amplitude equilibrium configurations, as the ergodicity is broken in many distinct states.

Four different photonic regimes are found varying  parameters ${\cal P}$, $\alpha$, $\alpha_0$ and $R_J$:
%\vspace{-0.15cm}
\begin{itemize}
 \item \emph{Incoherent Wave (IW):}
        replica symmetric solution with all order parameters equal to zero. 
        The modes oscillate incoherently and the light is emitted in the form of a continuous wave. 
        At low pumping this is the only solution. 
       It corresponds to the paramagnetic phase in spin models.
 \item \emph{Phase Locking Wave (PLW):}
       all order parameters vanish but the diagonal elements of $R_{\sfa \sfb} $, so that $ \langle \sigma^2 \rangle \neq \langle \tau^2 \rangle $ in this phase;
       this signals a partial locking of the mode phases.
       This regime occurs in the region of the phase space intermediate between the IW and RL regimes.
   %   if $\xi_4 > 0$ ($R_J>0$).
       This thermodynamic phase has not equivalent in other spin models, since 
       it follows directly from the peculiar nature of the degrees of freedom, that consist of both a phase and a magnitude,
       a combination of the well studied XY and spherical spin models.
\item \emph{Mode Locking Laser (ML):}
       solution with $m\not=0$, with or without replica symmetry breaking.
       The modes oscillate coherently with the same phase and the light is emitted in form of short pulses. % optical 
       It is the only regime at high pumping when the degree of disorder $R_J$ is small enough.
       It corresponds to the ferromagnetic phase in spin models.
 \item \emph{Random Laser (RL):}
       the modes do not oscillate coherently trivially (i.e., all with the same phase), so that $m=0$, 
       but % $R_{aa} = \langle \sigma^2 \rangle - \langle \tau^2 \rangle \neq 0$ implying a phase coherence
       %and 
       the overlap matrices $Q_{a b}$ and $R_{a b}$ have a nontrivial structure.
       It is the only phase in the high pumping limit for large degree of disorder $R_J$.
       It corresponds to the spin glass phase in spin models. % with disordered interactions. 
\end{itemize}
%\vspace{-0.15cm}
%[      We stress that the locking in the two degrees of freedom of each 
%       complex amplitude does not happen concurrently as the
%        pumping is increased in presence of (even a small amount of) 
%        disorder.  Mode phases lock first, in what we call the PLW regime. ]

The effect of the cavity leakages can be contemplated varying 
the strength of nonlinearity $\alpha$ and $\alpha_0$.
In particular, different replica symmetry breaking solutions RL regimes are possible
varying $\alpha$.
Consider, e.g., the situation shown in Fig. \ref{fig:RL1_a} of a large degree of disorder $R_J$,
so that $J_0 \ll J$ and the system is highly frustrated (the scenario is stable till the degree of disorder is so low that the ML phase appears).
For high $\alpha$ (closed cavity limit), the RL regime is 1RSB.
In this case the transition from PLW to RL is a random first order transition (RFOT) 
with the phenomenology typical of glass transition \cite{KirThiWol89}:
a jump is present in the order parameters $Q$ and $R$ but the internal energy remains continuous.
For $\alpha < \alpha_{\rm nl} \simeq (3-1.76382 \epsilon) (3-1.03703 \epsilon^2)^{-1} $, 
a continuous part must be considered in the distribution of the values for the elements of the overlap matrices.
The stable thermodynamic phase is, then,
first 1-FRSB and eventually a full replica symmetry breaking (FRSB) state \cite{CriLeu13}.  
In realistic optical systems the
$2$-body interaction is usually not dominant above the lasing threshold,
however,  there can be systems where the
damping due to the openness of the cavity is strong enough to compete
with the non-linearity and a FRSB state may emerge. In this case the transition turns out to be continuous in the
order parameters: the overlap is zero at the critical point and grows continuously as the power is increased above threshold.

\begin{figure}
\begin{center}
\subfigure{
\resizebox*{7.5cm}{!}{\includegraphics{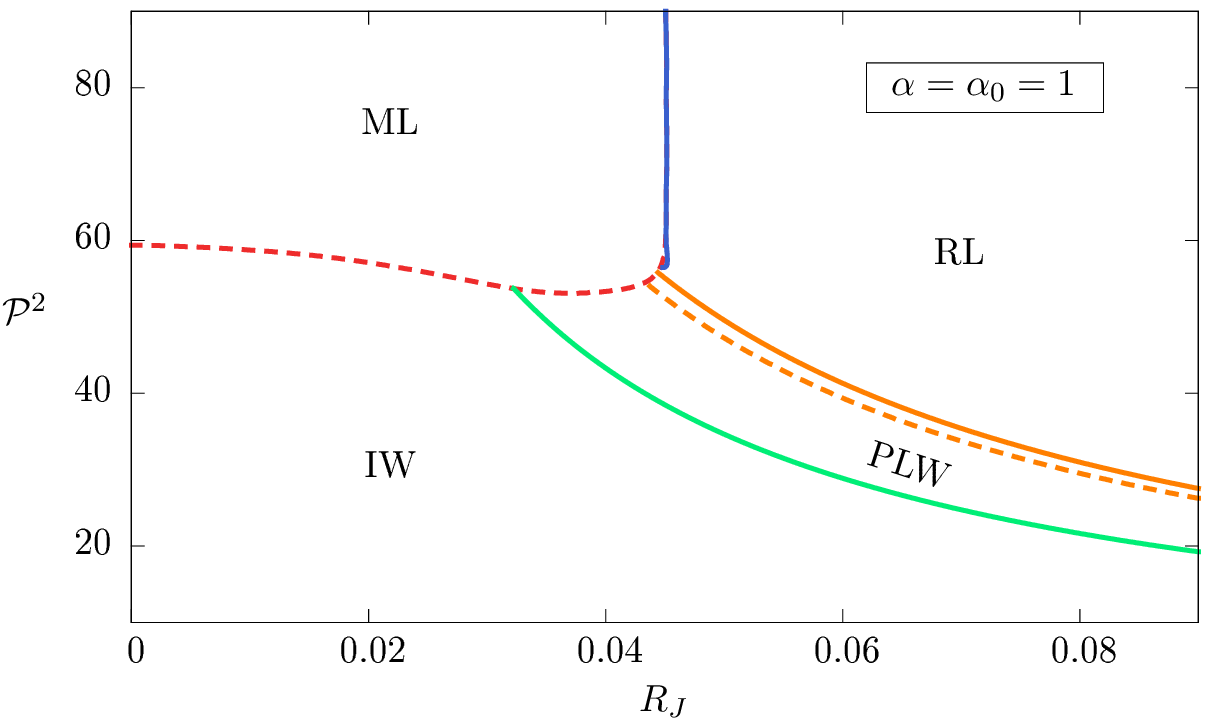}}}\hspace{30pt}
\subfigure{
\resizebox*{7.5cm}{!}{\includegraphics{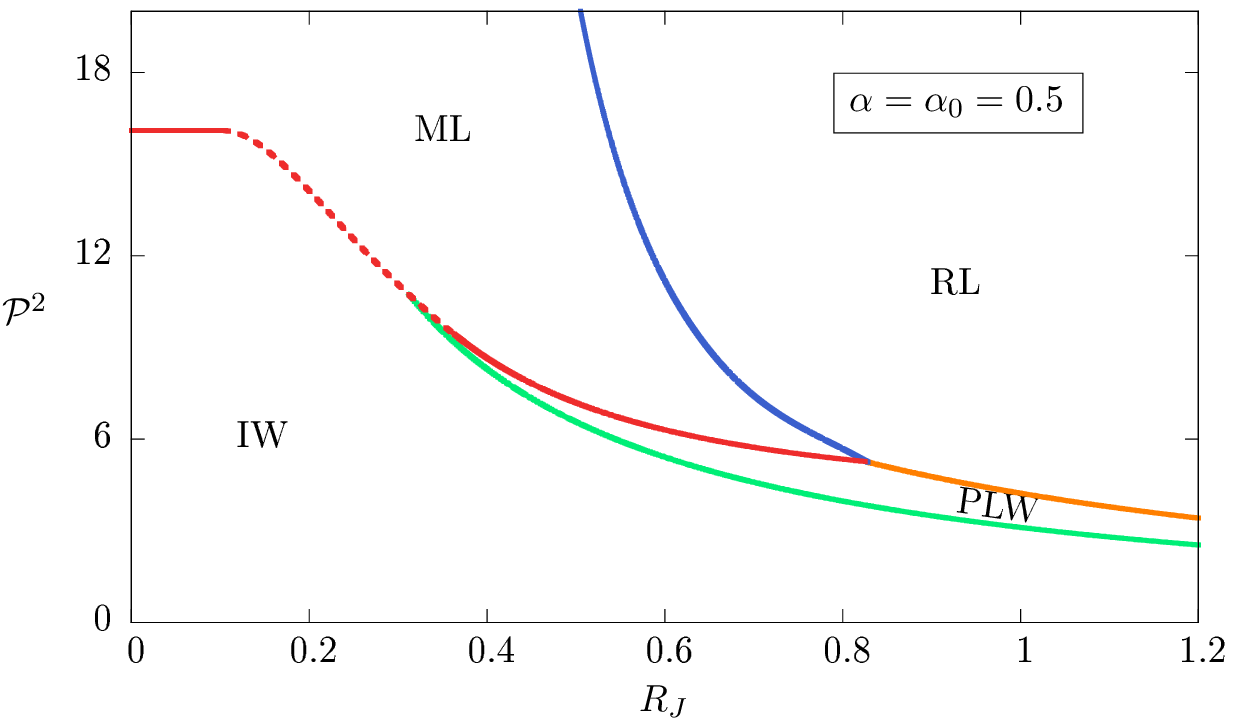}}}
\caption{ \label{fig:RL2}
Phase diagram for $\epsilon=1$ and $\alpha=\alpha_0=1$ corresponding to a %completely 
closed cavity (left) and $\alpha=\alpha_0=0.5$ corresponding to a highly opened cavity (right).
The solid (dashed) line between IW and ML corresponds to a continuous (discontinuous) transition.
The dashed line between PLW and RL in (a) indicates the dynamic transition line with finite complexity.
} 
\end{center}
\end{figure}

The whole phase diagram in terms of the four optical regimes is shown
in Fig. \ref{fig:RL1_b}, and two sections for in Fig. \ref{fig:RL2}, for every degree of disorder and $\alpha=\alpha_0$.
%Two typical phase diagrams are shown in Fig. \ref{fig:RL2}. % we show the phase diagram for $\alpha=\alpha_0=1$ (closed cavity) and $\alpha=\alpha_0=0.5$.
Consider the common experimental situation with an
increasing pumping rate $\mathcal{P}$ at fixed $R_J$,
$\alpha$ and $\alpha_0$. 
Then:
%\vspace{-0.15cm}
\begin{itemize}
\item for $R_J$ not too large a transition between the IW and the  ML regimes
    is observed increasing the pumping. 
    The transition is robust with respect to the introduction of small disorder;
\item for systems with intermediate disorder, the high pump regime 
remains the ordered ML regime, 
but the intermediate, partially coherent, PLW regime
 appears between ML and IW;
\item 
for large $R_J$, a further transition from ML to RL is observed at high ${\cal P}$.
Moreover, if $R_J$ exceeds a threshold the ML disappears and the only high pumping phase remains the RL.
\end{itemize}
%\vspace{-0.15cm}
%
%This scenario is rather  
%general and remains valid for different values of $\alpha$ and $\alpha_0$ \cite{ACCL}.
%
In general, the value of $\alpha_0$ affects the transition toward the ML
regime: for high $\alpha_0$ the transition is discontinuous.  
On the contrary, if $\alpha_0$ is low, there are regions in the phase diagram where 
the transition is continuous.
The value of $\alpha$ controls the transition to the RL regime.
For $\alpha > \alpha_{\text{nl}}$ ( $ \simeq 0.6297 $ for $\epsilon=1$)
%\begin{equation}
%\alpha > \alpha_{\text{nl}} =\frac{3-1.76382 \epsilon}{3-1.03703
%\epsilon^2}
%\end{equation}
% (in this work $\epsilon=1$ and $\alpha_{\text{nl}}= 0.6297\ldots$) 
the transition is toward a RL % phase %with a 1RSB structure 
via a RFOT, with a region of finite complexity antecedent the transition (cf. also Fig. \ref{fig:RL1_a})
where the photonic glass has an exponential number
of metastable states corresponding to different mode locking processes. % in the RL. 
%that is a continuous transition in the energy.
%Indeed, as disorder is introduced, already the standard mode locking lasing transition can become continuous, cf. Fig. \ref{fig:RL2}b,
%besides the random lasing transition. This difference can be experimentally tested.

\begin{figure}[t!]
\begin{center}
\includegraphics[width= 0.85 \columnwidth ]{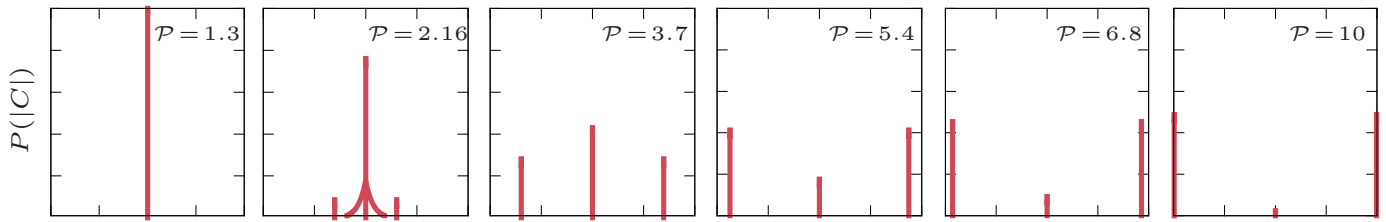}
\caption{
Probability distributions of the $\mathcal{C}$ overlap Eq. \eqref{eq:prescription} for $\alpha=\alpha_0=0.4$ and  $R_J = 1.1$.
Vertical lines represent Dirac deltas, whose height is the probability of the argument value. 
Different regimes are displayed from left to right: the first panel correspond to IW (the same distribution is obtained for PLW),
the second to RL(1-FRSB) and the last three to RL(1RSB) for increasing pumping.
}
\label{fig:RL3}
\end{center}
\end{figure}

As we aforementioned, the measure of the phase correlations, required not only by Eq. \eqref{eq:corrPhases}
but also for the static order parameters $Q$ and $R$ Eq. \eqref{eq:def_QR},
is hard because of the low intensity of RL emission
and has not been achieved so far.
Only data for the intensity spectrum are usually available.
We note that the lack of a direct experimental measure of the overlap
probability distribution is common to the original 
systems for which replica theory was first
developed, i.e., spin-glasses \cite{Edwards75,Sherrington75}, and
also to structural glasses, one of the fields of major application of
the theory \cite{MezPar99a,ParZam10,Caltagirone12,Franz12,Charbonneau14}.
The inclusion of the amplitude as degree of freedom in the previous analysis 
paves the way to a test directly from intensity data.
Indeed an experimental validation of a random laser-glass
connection, and, specifically, of the presence of RSB predicted by the
theory, has recently been put forward in Ref. \cite{Ghofraniha15}, 
measuring the overlap between intensity fluctuations
of the spectral emission. 
In the model Eq. \eqref{eq:langevin_slow}, it is possible to define 
the intensity fluctuations overlap matrix 
\begin{align}
\label{eq:prescription}
 \mathcal{C}_{\sfa \sfb} \equiv & \frac{1}{8 N \epsilon^2}
 \sum_{k=1}^N \left[
  \langle |a^{\sfa}_k|^2  |a^{\sfb}_k|^2 \rangle
 - \langle |a^{\sfa}_k|^2 \rangle \langle |a^{\sfb}_k|^2 \rangle \right] 
\end{align}
defined in the dominion $[0,1]$. In the following we consider the symmetrized 
probability distribution $P(|\mathcal{C}|)$, without any loss of generality.
It can be shown \cite{AntCriLeu15} that for $m=0$ it holds simply $\mathcal{C}_{\sfa \sfb} = Q_{\sfa \sfb}^2$,
so that the intensity fluctuations overlap is nontrivial
in the RL regime and the replica symmetry breaking can be studied from intensity spectrum data.
We note also that instead, for low disorder, at the IW-ML transition $\mathcal{C}_{\sfa \sfb}$
does not change at the transition \cite{AntCriLeu15}.
A typical scenario for the overlap $\mathcal{C}$ in an open cavity for high disorder is shown in Fig. \ref{fig:RL3}:
at the RL transition the distribution becomes nontrivial with more than one value possible, 
meaning that the replica symmetry is broken.
The behavior shown in Fig. \ref{fig:RL3} is qualitatively reproduced by the experimental data of Ref. \cite{Ghofraniha15},
where the overlap is measured between the intensity fluctuations of different shots (assumed as \emph{real replicas})
in the regime where the shot-to-shot fluctuations are observed.
These results support, then, the interpretation of the observations as the first direct experimental 
evidence of replica symmetry breaking.

%%%%%%%%%%%%%%%%%%%%%%%%%%%%%%%%%%%%%%%%%%%%%%%%
\section{Conclusions}

In this paper we have presented the main results of the application of statistical mechanics to  nonlinear photonic systems, both in standard laser 
systems with optical cavities and in cavity-less random media.
For standard lasers the outcome have been quite wide and fruitful,
providing a clear and universal description of key phenomena like 
active and passive mode locking and, in particular, the role of the non-linear  frequency matching for the
onset of pulsed emission and the gain narrowing  in ultra-fast multimode lasers.

We have, further, pointed out the crucial role that the statistical mechanic approach can play in the study of  photonic systems in disordered media, like random lasers. In the latter case, the techniques developed for complex systems turn out to be a valid theoretical tool to consider nonperturbatively the concurrent presence of disorder and  nonlinearity including many aspects not yet established in theoretical optics.
In order to have a complete theory
for light-wave propagation and amplification in disordered active media,
e.g., 
the structure of the lasing modes in an open cavity for realistic systems above the lasing threshold has yet to be understood.
The presence of a complex structure of modes in such systems was revealed already in the first experiments,
by the presence of
narrow resonances in the emission spectrum, emerging as the pumping was increased \cite{Froetal99},  
and supported by later studies of photon statistics \cite{Poletal01,Cao2001}.
In the course of the last decade, random lasing has been observed in many different kinds of disordered
materials, like polymer films \cite{Poletal01}, porous materials \cite{vanderMolen07}, powders \cite{Cao99}, 
ceramics \cite{ceramicRL1,ceramicRL2}, clusters \cite{clustersRL}, colloidal solutions of 
nanoparticles \cite{colloidalRL}, and it is now generally regarded as a universal property of optically active disordered structures.
However, the physical mechanism underlying the constitution 
of long-living laser modes is, nonetheless, not yet known.
Some attempts to investigate their formation have been made on the basis of the \emph{light localization} \cite{Sajeev84,Anderson85,Lagendijk86},
which, however, is very difficult to accomplish in optics  \cite{Abretal79,SkiSok14}.
In most materials, indeed, the intensity is spread throughout the sample and the modes appear to be extended to the whole system.
An alternative explanation is based on the presence of rare long light paths in the material \cite{Mujetal07}:
the spikes in the emission spectrum are associated with ``lucky photons'' spontaneously emitted
in such long paths and acquiring a huge gain, without the need for strong localization. 
Depending on the specific setting, it is expected that
both localized and extended modes can, in principle, participate in amplification by stimulated emission and which kind of modes is dominant
depends, in general, on the delicate balance between gain saturation, scattering strength and
amplification length  \cite{Fallert09}.

To discriminate universal and particular characteristics in such composite situation is not an easy task
and a statistical mechanics approach can help to achieve a classification based on fundamental properties.
We have reported the main results in this context that, up to now, are limited
to mean field theories of effective Hamiltonian models of the mode interaction.
In particular, we have described 
the general phase diagrams, ranging from ordered closed
cavities to disordered open cavities,
for the most general model
that includes the presence of leakages and the amplitude dynamics.
An experimental test for replica symmetry breaking at the random lasing threshold can be put forward within this theory:
the replica symmetry breaking in the intensity fluctuation overlap is shown to be equivalent to the one occurring 
in the standard Parisi overlap (the complex amplitude overlap), allowing for a verification of Replica Symmetry Breaking occurrence 
directly from shot-to-shot fluctuations of emission spectra  \cite{Ghofraniha15}. The first direct experimental evidence, to our knowledge.
Our results also hint that a classification 
can be established between glassy and non-glassy random lasers, 
depending on the distribution of the disordered nonlinear
 mode-couplings induced by the 
microscopic structure of wave scatterers and the properties of the optically active material in the candidate random laser compound.

There are several possible developments to the models studied so far for random lasing. 
%beyond mean field Hamiltonian approach.
The most immediate progress would be to go beyond the mean field theory and
consider the role of the frequency matching condition 
%(as implemented for the ordered case, cf. section \ref{sec:ordered})
and the finite spatial extension of the modes.
As discussed for the case of standard lasers, this is possible employing Monte Carlo simulations
or the cavity method on suited graphs.
In this respect, we note that the frequency matching condition has the same structure 
present in the Golay-Bernasconi model \cite{Golay82,Bernasconi87}, with discrete Boleean variables,
a model without quenched disorder but with a complex landscape 
and a glassy behavior and that has been studied via the replica method \cite{MarParRit94a,MarParRit94b}.
%The difference is that the Golay-Bernasconi Hamiltonian is defined with an opposite sign 
%with respect to the photonic one, cf. Eq. \eqref{eq:langevin_slow},
%so the approach used in that case is not directly useful to study 
%the mode locking regime 
%\footnote{
%indeed in the Golay-Bernasconi model one is usually interested
%in the opposite limit with respect the mode locking, corresponding 
%to sequences with low autocorrelation and a high \emph{merit factor}.
%}.
The most feasible alternative to go beyond mean field is to use numerical techniques like
Monte Carlo simulations, as accomplished for the ordered case (cf. section \ref{sec:ordered}).
In this way one can also study finite size effects that may be relevant in 
connection with experiments. % \cite{Antenucci_inprogress}.

An important counterpart of the analysis here reported is to directly study
the dynamics, in its general evolution described by the Langevin equations 
for the electromagnetic modes, without the requirement of equilibrium and Hamiltonian dynamics.
In this case, in particular, it would be possible to investigate 
the role of dispersion and of the gain dynamics preceding  saturation and the onset of the stationary lasing regime.

A preeminent issue remains the fundamental structure of the lasing modes 
in cavity-less and disordered media and the shape of their effective interaction.
A first aspect is that the interactions are not, in general, independent from each other,
as assumed in the mean field approximation for large number of modes 
(possible correlations are expected to decay with the size of the system).
To have access to the values of such interactions in realistic settings
is, though, not an easy task.
We note that, in such an intent, statistical mechanics can again play an important role
providing the statistical inference techniques to reconstruct the interactions from
configuration data (see, e.g. Ref. \cite{Tyagietal15}).

The benefits obtained so far through statistical physics in the study of laser physics 
support the use of such framework in more setups in future.
One peculiar and relevant case might, e.g., be the study of free-electron lasers,
where the lasing is obtained by high-speed electrons moving in an inhomogeneous magnetic structure
\cite{Milton15062001,ZhiKwa07}.

Concluding, we remind that an early motivation 
for the development of the theory of complex systems was the analysis of
\emph{``certain alloys of ferromagnets and conductors, such as AuFe or CuMg, known as Spin Glasses,
[...] intriguing and perplexing as no others in the history of solid state physics''}
\cite{bolthausen2007random}. 
The theory actually found quickly many areas of application such as structural glasses, neural networks, 
financial markets, granular materials,
the immune system, road traffic, flocking in birds or fish, human social communities.
After all the above mentioned results that have been obtained in the last decades, 
we feel that also optics can now be entirely added to the list.

\section*{Funding}
The research leading to these results has received funding from 
the People Programme (Marie Curie Actions) 
of the European Union's Seventh Framework Programme FP7/2007-2013/ under REA grant agreement n¡ 290038, NETADIS project,
from the European Research Council through ERC grant agreement no. 247328 - CriPheRaSy project - 
and from the Italian MIUR under the
Basic Research Investigation Fund FIRB2008 program, grant No. RBFR08M3P4, and under the PRIN2010 program, 
grant code 2010HXAW77-008.

\bibliography{FabBib_v3}

\end{document}